\documentclass[aps,prd,12pt]{article}
\usepackage{graphicx}
\usepackage{graphics,epsfig}
\usepackage{amssymb,amsmath,bm}
\usepackage{hyperref} 

\usepackage[T1]{fontenc}
\usepackage[utf8]{inputenc}
\usepackage{authblk}

\usepackage{slashed}
\usepackage{pifont}

\renewcommand{\thefootnote}{\fnsymbol{footnote}}
\newcommand{\ubar}[1]{\overline{U}_{#1}}
\newcommand{\vbar}[1]{\overline{V}_{#1}}
\newcommand{\tvec}[1]{\vec{#1}_\bot}
\def\eq#1{{Eq.~(\ref{#1})}}
\def\fig#1{{Fig.~\ref{#1}}}

\def\ru1{\rule[-0.4truecm]{0mm}{1truecm}}
\title{\vspace*{1cm} \Large \bf Single-Spin Asymmetries in
  Semi-inclusive Deep Inelastic Scattering and Drell-Yan Processes}

\author[a]{Stanley J. Brodsky} 
\author[b]{Dae Sung Hwang}
\author[c]{Yuri V. Kovchegov}
\author[d]{~\\ Ivan Schmidt}
\author[c]{Matthew D. Sievert}

\affil[a]{\small SLAC National Accelerator Laboratory, Stanford
  University, Stanford, CA 94309, USA}

\affil[b]{\small Department of Physics, Sejong University, Seoul 143--747,
  South Korea}

\affil[c]{\small Department of Physics, The Ohio State University, Columbus,
  OH 43210, USA}

\affil[d]{\small Departamento de F\'\i sica, Universidad T\'ecnica
  Federico Santa Mar\'\i a, Casilla 110-V, Valpara\'\i so, Chile}

\date{April 2013}

\begin{document}

\maketitle

\thispagestyle{empty}

\begin{abstract}
  We examine in detail the diagrammatic mechanisms which provide the
  change of sign between the single transverse spin asymmetries
  measured in semi-inclusive deep inelastic scattering (SIDIS) and in
  the Drell-Yan process (DY). This asymmetry is known to arise due to
  the transverse spin dependence of the target proton combined with a
  $T$-odd complex phase. Using the discrete symmetry properties of
  transverse spinors, we show that the required complex phase
  originates in the denominators of rescattering diagrams and their
  respective cuts.  For simplicity, we work in a model where the
  proton consists of a valence quark and a scalar diquark. We then
  show that the phases generated in SIDIS and in DY originate from
  distinctly different cuts in the amplitudes, which at first appears
  to obscure the relationship between the single-spin asymmetries in
  the two processes. Nevertheless, further analysis demonstrates that
  the contributions of these cuts are identical in the leading-twist
  Bjorken kinematics considered, resulting in the standard sign-flip
  relation between the Sivers functions in SIDIS and DY.  Physically,
  this fundamental, but yet untested, prediction occurs because the
  Sivers effect in the Drell-Yan reaction is modified by the
  initial-state ``lensing'' interactions of the annihilating antiquark,
  in contrast to the final-state lensing which produces the Sivers
  effect in deep inelastic scattering.
\end{abstract}

~\\
\centerline{PACS numbers: 12.38.Bx, 13.88.+e}

\begin{flushright}
\vspace{-20.5cm}
{\normalfont SLAC-PUB-15408
\\}
\end{flushright}
\thispagestyle{empty}

\setcounter{page}{0}

\newpage

\setcounter{footnote}{0}
\renewcommand{\thefootnote}{\arabic{footnote}}


\section{ Introduction}

\subsection{Factorization and Final State Interactions}

The factorization picture of leading twist-perturbative QCD has played
a guiding role in virtually all aspects of hadron physics
phenomenology.  In the case of inclusive reactions such as
hadroproduction at large transverse momentum $p + p \to H+
X$~\cite{Berman:1971xz,Arleo:2009ch}, the parton model for
perturbative quantum chromodynamics (pQCD) predicts that the cross
section at leading order in the transverse momentum $p_T$ can be
computed by convoluting the perturbatively calculable hard subprocess
quark and gluon cross section with the process-independent structure
functions of the colliding hadrons and the final-state quark or gluon
fragmentation functions.  The resulting leading-twist cross section
$E_H {d\sigma/d^3 p_H}( p p \to HX)$ scales as $1/ p^4_T,$ modulo the
DGLAP scaling violations derived from the logarithmic evolution of the
structure functions and fragmentation distributions, as well as the
running of the QCD coupling appearing in the hard-scattering
subprocess matrix element.

The effects of final-state interactions of the scattered quark in deep
inelastic scattering are characterized by a Wilson line. Such effects
have traditionally been assumed to either give an inconsequential
phase factor or power-law suppressed corrections in hard pQCD
reactions.  However, this expectation is only true for sufficiently
inclusive cross sections.  For example, consider semi-inclusive deep
inelastic lepton scattering (SIDIS) on a transversely polarized target
$\ell + p^\uparrow \to H + \ell' + X$. (For a review see
\cite{Brodsky:2011fa}.) In this case the final-state gluonic
interactions of the scattered quark lead to a pseudo-$T$-odd non-zero
spin correlation of the lepton-quark scattering plane with the
polarization of the target proton~\cite{Brodsky:2002cx} which is not
power-law suppressed with increasing virtuality of the photon $Q^2$;
i.e., it Bjorken-scales.  This asymmetry is made experimentally
explicit by studying the triple product of vectors $\vec S_P \cdot
(\vec q \times {\vec p}_H)$, where ${\vec p}_H$ is the momentum of the
hadron fragmented from the struck quark jet.  Similar correlations can
be found from target spin asymmetries, due to multi-photon
exchanges~\cite{Metz:2012ui}.

A crucial fact is that the leading-twist ``Sivers
effect''~\cite{Sivers:1989cc,Sivers:1990fh} is non-universal in the
sense that pQCD predicts an opposite-sign correlation in Drell-Yan
reactions relative to semi-inclusive deep inelastic
scattering~\cite{Collins:2002kn,Brodsky:2002rv}.  This fundamental,
but yet untested, prediction occurs because the Sivers effect in the
Drell-Yan reaction is modified by the initial-state ``lensing''
interactions of the annihilating antiquark, in contrast to the
final-state lensing which produces the Sivers effect in deep inelastic
scattering.

\begin{figure}
 \begin{center}
\includegraphics[width=16cm]{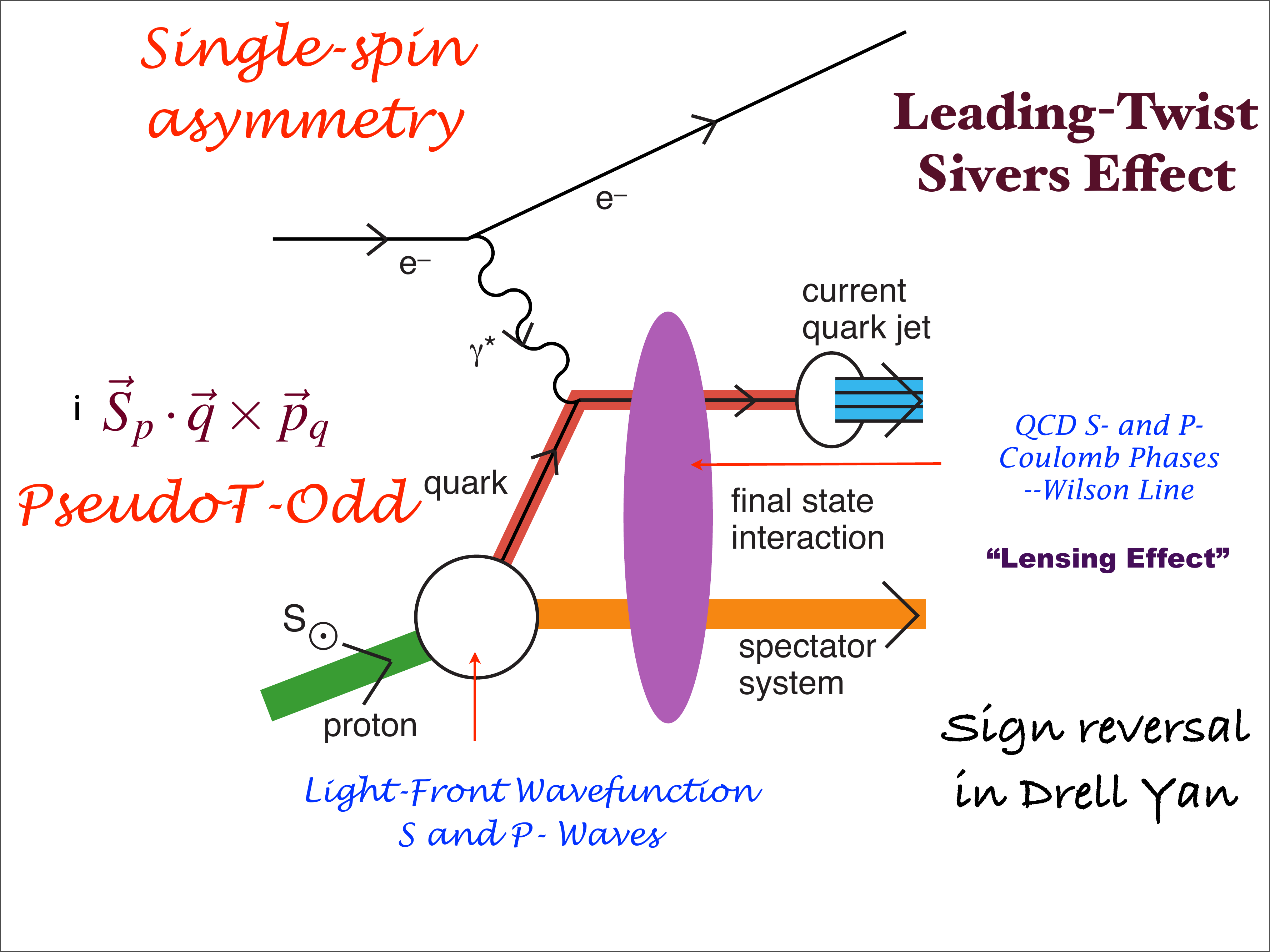}
\end{center}
\caption{Origin of the Sivers single-spin asymmetry in semi-inclusive
  deep inelastic lepton scattering.}
\label{Sivers}  
\end{figure}

The calculation of the Sivers single-spin asymmetry in deep inelastic
lepton scattering in QCD is illustrated schematically in
Fig.~\ref{Sivers}. Although the Coulomb phase for a given partial wave
is infinite, the interference of Coulomb phases arising from different
partial waves leads to observable effects. The analysis requires two
different orbital angular momentum components: $S$-wave with the
quark-spin parallel to the proton spin and $P$-wave for the quark with
anti-parallel spin; the difference between the final-state ``Coulomb''
phases leads to a $\vec S \cdot (\vec q \times \vec p)$ correlation of
the proton's spin with the virtual photon-to-quark production
plane~\cite{Brodsky:2002cx}.  Thus, as it is clear from its QED
analog, the final-state gluonic interactions of the scattered quark
lead to a pseudo--$T$-odd non-zero spin correlation of the
lepton-quark scattering plane with the polarization of the target
proton~\cite{Brodsky:2002cx}.  The effect is pseudo-T odd due to the
imaginary phase generated by the cut of the near-on-shell intermediate
state.

The $S$- and $P$-wave proton wavefunctions also appear in the
calculation of the Pauli form factor quark-by-quark. Thus one can
correlate the Sivers asymmetry for each struck quark with the
anomalous magnetic moment of the proton carried by that
quark~\cite{Lu:2006kt}, leading to the prediction that the Sivers
effect is larger for positive pions as seen by the HERMES experiment
at DESY~\cite{Airapetian:2004tw}, the COMPASS
experiment~\cite{Bradamante:2011xu,Alekseev:2010rw,Bradamante:2009zz,Adolph:2012sp}
at CERN, and CLAS at Jefferson
Laboratory~\cite{Avakian:2010ae,Gao:2010av}.

The final-state interactions of the produced quark with its comoving
spectators in SIDIS also produces a final-state $T$-odd polarization
correlation -- the ``Collins effect'', in which the Collins
fragmentation function, instead of the Sivers distribution function,
generates the spin asymmetry~\cite{Collins:1992kk,Boer:1997nt}. This
can be measured without beam polarization by measuring the correlation
of the polarization of a hadron such as the $\Lambda$ baryon with the
quark-jet production
plane~\cite{Anselmino:2000vs,Gustafson:1992iq,Ma:2000uv,Ma:2000uu}. Analogous
spin effects occur in reactions in QED due to the rescattering via
final-state Coulomb interactions.

In principle, the physics of the ``lensing dynamics'' or Wilson-line
physics~\cite{Burkardt:2003uw,Brodsky:2010vs} underlying the Sivers
effect involves nonperturbative quark-quark interactions at small
momentum transfer, not the hard scale $Q^2$ of the virtuality of the
photon.  These considerations have thus led to a reappraisal of the
range of validity of the standard factorization ansatz.  As noted by
Collins and Qiu~\cite{Collins:2007nk}, the traditional factorization
formalism of perturbative QCD fails in detail for many hard inclusive
reactions because of initial- and final-state lensing interactions.
For example, if both the quark and antiquark in the Drell-Yan
subprocess $q + \bar q \to \mu^+ + \mu^-$ interact with the spectators
of the other hadron, then one predicts a $\cos 2\phi \, \sin^2 \theta$
planar correlation in unpolarized Drell-Yan
reactions~\cite{Boer:2002ju}. (Here $\phi$ is the angle between the
muon plane and the plane of the incident hadrons in the lepton pair
center of mass frame, while $\theta$ is the angle between the momentum
of one of the muons and the line of flight of partons in the same
frame~\cite{Boer:2002ju,Falciano:1986wk}.) This ``double Boer-Mulders
effect'' \cite{Boer:1997nt} can account for the anomalously large
$\cos 2 \phi$ correlation and the corresponding
violation~\cite{Boer:2002ju, Boer:1999mm} of the Lam-Tung relation
\cite{Lam:1978pu,Lam:1980uc} for Drell-Yan processes observed by the
NA10 collaboration~\cite{Falciano:1986wk}, and the azimuthal angle
dependence of di-jet production in unpolarized hadron
scattering~\cite{Lu:2008}. Such effects again point to the importance
of the corrections from initial and final-state interactions of the
hard-scattering constituents, which are not included in the standard
pQCD factorization formalism.

The final-state interactions of the struck quark with the target
spectators~\cite{Brodsky:2002ue} also lead to diffractive events in
deep inelastic scattering (DIS) at leading twist, such as $\ell + p
\to \ell' + p' + X$, where the proton remains intact and isolated in
rapidity; in fact, approximately 10 \% of the deep inelastic
lepton-proton scattering events observed at HERA are
diffractive~\cite{Adloff:1997sc, Breitweg:1998gc}. The presence of a
rapidity gap between the target and diffractive system requires that
the target remnant emerges in a color-singlet state; this is made
possible in any gauge by the soft rescattering incorporated in the
Wilson line or by augmented light-front wavefunctions.



\subsection{The Sign of the Sivers Effect}

\begin{figure}
\centering
\includegraphics[width=0.73 \textwidth]{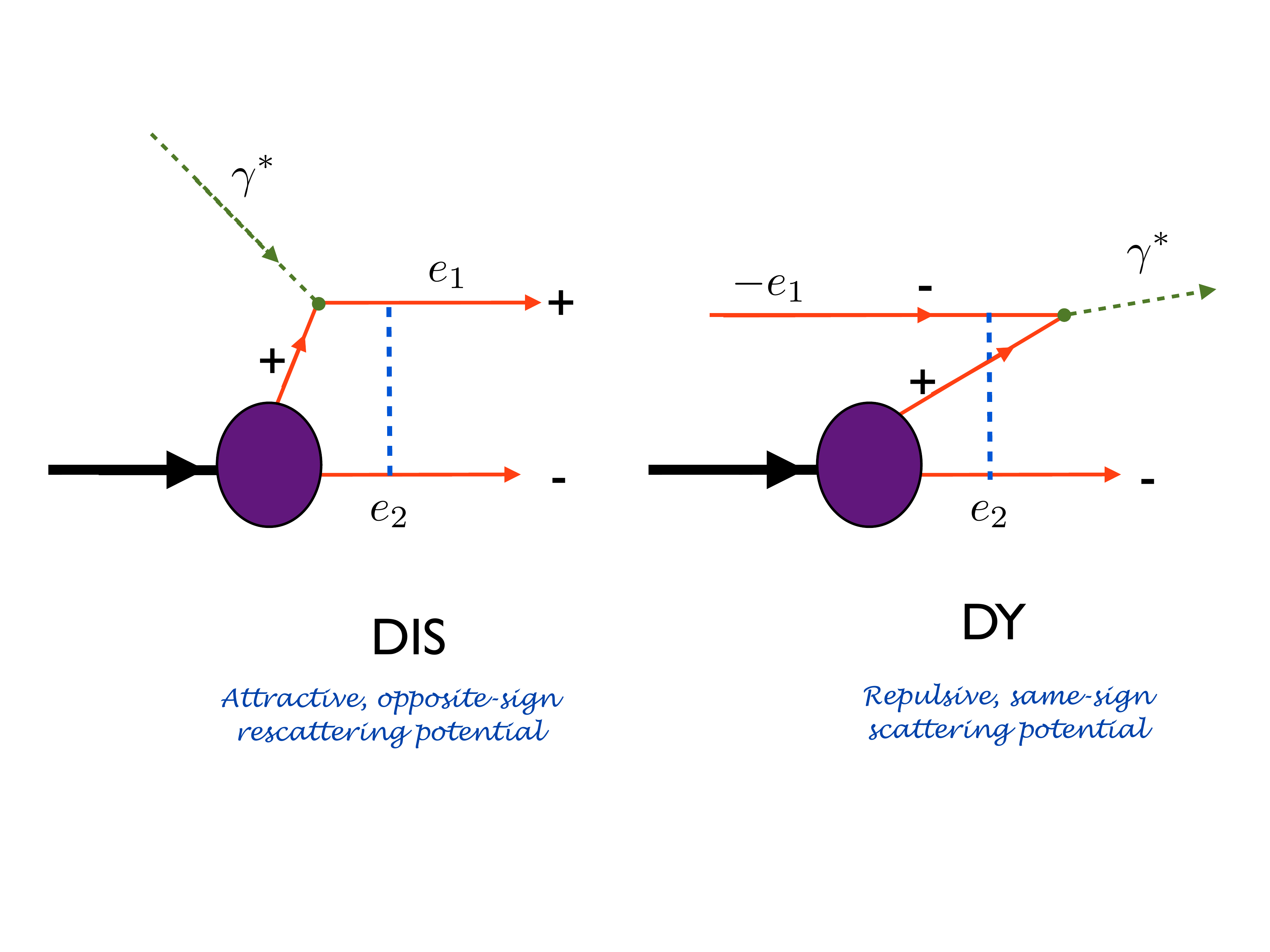}
\caption{The underlying physics of the sign change of the lensing
  interactions in SIDIS and DY reactions for QED. Dashed lines denote
  the photon propagators. The final-state interaction in SIDIS is
  attractive whereas the initial state interaction is repulsive in
  DY. }
\label{relsign}
\end{figure}

As we have emphasized in the above, the sign reversal predicted by QCD
between the Sivers functions
\cite{Sivers:1989cc,Sivers:1990fh,Brodsky:2002rv} in semi-inclusive
deep inelastic scattering (SIDIS) and in the Drell-Yan process (DY)
\cite{Collins:2002kn,Brodsky:2002rv} is in contrast to the standard
expectations of collinear pQCD/parton model factorization.  This
reversal results in different signs predicted for the single
transverse spin asymmetry (SSA) in SIDIS on a transversely polarized
proton versus that in the DY process involving a transversely
polarized hadron. The sign difference of the Sivers functions is a
consequence of their being pseudo $T$-odd, with the SIDIS Sivers
function related to the DY Sivers function by a time-reversal
transformation \cite{Collins:2002kn,Belitsky:2002sm,Boer:2003cm}.  The
essential physics of the sign change is illustrated for QED in
\fig{relsign}. The final-state interaction in SIDIS is attractive
whereas the initial-state interaction is repulsive in DY.  However,
the detailed analysis involves many subtle aspects due to intermediate
Glauber-type cuts related to unitarity.  These considerations are
worked out in detail in this article.

In general, the generation of a SSA requires two ingredients: (i) the
dependence of the cross section on the transverse polarization, and
(ii) a relative complex phase between the amplitudes corresponding to
different orbital angular momenta
\cite{Jianwei,Qiu:1991pp,Qiu:1998ia}. In the original argument given
by Collins \cite{Collins:2002kn} the sign flip between SSA in SIDIS
and DY arises due to a reversal of the direction of the Wilson line in
the definition of the Sivers function under the application of
$T$-reversal. It is interesting and important to clarify the relation
between this time-reversal argument and the diagrammatic conditions
(i) and (ii) stated above. While it is well-known that the propagation
of a high energy quark or gluon can be approximated by a Wilson line
along the corresponding light cone direction, the connection between
the Wilson lines and Feynman diagrams can be non-trivial
\cite{Mueller:2012bn}, especially in the SSA case, where we are
interested in the imaginary part of the scattering amplitude due to
the condition (ii) above. Establishing a clear connection in the
Feynman diagram language between the spin asymmetries in SIDIS and DY
will greatly aid future (model) diagrammatic calculations for the
processes.

Although the time-reversal symmetry argument allows one to confidently
predict the sign flip between the Sivers functions in the two
processes, a detailed diagrammatic model calculation demonstrating the
origin of this sign reversal appears to be lacking in the literature. A
model calculation for the SSA in SIDIS was first constructed in
\cite{Brodsky:2002rv} by some of the authors; however, an analogous
calculation for DY outlined in \cite{Brodsky:2002cx} assumed that the
relative phase arises in DY due to putting the same propagators as in
SIDIS on mass shell. As we will show below, one can think of the
relative phase from the condition (ii) as a cut through an amplitude
(or a complex conjugate amplitude), in addition to the standard
final-state cut one obtains when the scattering amplitude is
squared. In terms of the cuts, the calculation of
\cite{Brodsky:2002cx} assumed that the cut generating SSA in SIDIS was
equivalent to the cut in DY. While this assumption allows one to obtain
the correct answer with the sign flip in the Sivers function between
SIDIS and DY, as we shall discuss below, it is not easy
to justify.  In fact, the explicit calculation of the SSA in the
SIDIS and DY processes given in this paper demonstrates that the cuts are, in fact,
different, as illustrated in Figures~\ref{DIS-DY2} and
\ref{DIS-DY1}. In Feynman diagram language this means that
different propagators are put on the mass shell in order to extract the asymmetry
in the two processes. In light-front perturbation theory (LFPTH)
\cite{Lepage:1980fj,Brodsky:1997de} language this corresponds to
putting different intermediate states (that is, intermediate states
involving different particles) on the $P^-$ LF energy shell.

One may then wonder how the simple sign-flip relation between the
Sivers functions in SIDIS and DY would arise from Feynman diagrams
with different cuts (or LFPTH diagrams with different intermediate
states put on energy shell). While further clarification of the
diagrammatic origin of the sign flip is left for future work, we
shall show here by performing an explicit calculation that in the limit of high
energy $s$ and high photon virtuality $Q^2$, the simple sign-flip
relationship is preserved for SIDIS and DY Sivers functions, despite
the different cuts in the diagrams in the two cases. We thus are able to construct
a diagrammatic description of the sign reversal.

The paper is structured as follows. In Section~\ref{sec:phase} we show
how the transverse polarization dependence always comes into the
amplitude squared with the imaginary factor of $i$. Since the cross
section contribution has to be real, this implies that, for the spin
asymmetry to be non-zero, another factor of $i$ has to be generated
elsewhere in the amplitude squared. We thus demonstrate the need for a
phase difference between the amplitude and the complex conjugate
amplitude contributing to the SSA \cite{Qiu:1991pp,Qiu:1998ia}.  Our
analysis provides an explicit verification of the analytical
properties of the complete amplitudes.

In Section~\ref{ModelFeynman} we perform a model calculation of the
spin asymmetry in the SIDIS and DY cases using a Feynman diagram
approach. We employ a model in which the proton is made out of a
valence quark and a scalar color-charged diquark: this model was used
previously in \cite{Brodsky:2002cx,Brodsky:2002rv}. Instead of
calculating the Sivers functions directly, we consider the $\gamma^* +
p^\uparrow \to q + X$ process in place of the SIDIS Sivers function,
and the ${\bar q} + p^\uparrow \to \gamma^* + X$ process in place of
the DY Sivers function. It can be shown that the relation between
these processes and the corresponding Sivers functions is the same in
both cases (see \cite{Boer:2002ju} for an example of how to establish
the connection): hence, the sign reversal of the SSAs in the $\gamma^* +
p^\uparrow \to q + X$ and ${\bar q} + p^\uparrow \to \gamma^* + X$
processes is equivalent to the sign flips in the corresponding Sivers
functions.

Performing an explicit calculation in Section~\ref{ModelFeynman} we
demonstrate that, in the framework of the model considered, the SSAs
in SIDIS and DY arise from different cuts, as shown in
Figs.~\ref{DIS-DY2} and \ref{DIS-DY1}. Nevertheless, the resulting
spin asymmetries only differ by a minus sign, in agreement with
the general arguments based on time-reversal anti-symmetry
\cite{Collins:2002kn,Belitsky:2002sm}. We demonstrate how the same
conclusion can be achieved in a LFPTH calculation in
Section~\ref{ModelLC}: while different LF energy denominators are put on
energy shell in SIDIS and DY, the resulting SSA in DY is simply a
negative of that in the SIDIS case. 
Section~\ref{sec:conclusions} summarizes our main results.


\section{Coupling of Transverse Spinors to the Complex Phase}
\label{sec:phase}

The discrete symmetries of the Dirac equation, $\mathrm{C, P}$ and
$\mathrm{T}$ provide strong constraints on the form of the spinors and
their matrix elements.  In particular, the $\mathrm{PT}$ properties of
transverse spinors bring in a unique coupling of the transverse spin
dependence of a matrix element to its real and imaginary parts.  This
gives rise to the conclusion that the single transverse spin asymmetry
requires the presence of a complex phase aside from the spinor matrix
elements themselves.

Below, we explicitly construct the transverse spinors and the
identities they satisfy corresponding to $\mathrm{C/P/T}$
transformations.  We demonstrate that, as a consequence, when two
spinor matrix elements are multiplied together (as in the numerator of
a scattering amplitude squared), the part which depends on the
transverse spin is pure imaginary.

We then define the asymmetry observable $A_N$ and use the knowledge of
the $\mathrm{C/P/T}$ constraints to identify which types of diagrams
can contribute to the asymmetry.  We find that only interference terms
can contribute to $A_N$, and furthermore only the imaginary part of
the interference term (aside from the spinor products) contributes.
We derive an expression for the spin-difference part of the amplitude
squared in terms of the imaginary part of the rest of the interference
term.


\subsection{Spinor Conventions and $\mathrm{C/P/T}$ Identities}

In this calculation we will work with the spinors defined in Ref.
\cite{Lepage:1980fj}.  These spinors correspond to the spin projection
along the $\pm z$ axis of a particle with mass $m$ and are expressed
in light-cone coordinates $p^\pm \equiv p^{(0)} \pm p^{(3)}$.  For
definiteness, working in the standard (Dirac) representation of the
Clifford algebra, this spinor basis is:
\begin{eqnarray}
\label{spinors1}
U_{+z} (p) &= \frac{1}{\sqrt{2p^+}} \left[ 
 \begin{array}{c} p^+ + m \\ p^{(1)} + i p^{(2)} \\ p^+ - m \\ p^{(1)} + i p^{(2)} \end{array}
 \right] \hspace{1cm}
U_{-z}(p) &= \frac{1}{\sqrt{2p^+}} \left[ 
 \begin{array}{c} -p^{(1)} + i p^{(2)} \\ p^+ + m \\ p^{(1)} - i p^{(2)} \\ -p^+ + m \end{array}
 \right]
\\ \nonumber
V_{+z} (p) &= \frac{1}{\sqrt{2p^+}} \left[ 
 \begin{array}{c} -p^{(1)} + i p^{(2)} \\ p^+ - m \\ p^{(1)} - i p^{(2)} \\ -p^+ - m \end{array}
 \right] \hspace{1cm}
V_{-z}(p) &= \frac{1}{\sqrt{2p^+}} \left[ 
 \begin{array}{c} p^+ - m \\ p^{(1)} + i p^{(2)} \\ p^+ + m \\ p^{(1)} + i p^{(2)} \end{array}
 \right] .
\end{eqnarray}

Like any spinor basis for solutions of the Dirac equation, these
spinors satisfy identities that embody the discrete $\mathrm{C}$,
$\mathrm{P}$, and $\mathrm{T}$ symmetries of the theory.  As can be
explicitly verified from (\ref{spinors1}), these spinors obey the
identities
\begin{eqnarray}
\label{CPT1}
\mathrm{C:} \hspace{1cm} & -i \gamma^2 V_{\pm z}^* (p) = U_{\pm z} (p) \\ \nonumber
\mathrm{PT:} \hspace{1cm} & \gamma^1 \gamma^3 \gamma^0 U_{\pm z}^* (p) = \mp U_{\mp z} (p) \\ \nonumber
                          & \gamma^1 \gamma^3 \gamma^0 V_{\pm z}^* (p) = \pm V_{\mp z} (p) \\ \nonumber
\mathrm{CPT:} \hspace{1cm} & U_{\pm z} (p) = \pm \gamma^5 V_{\mp z} (p) ,
\end{eqnarray}
where the final $\mathrm{CPT}$ identity combines the other two in a
compact form.

For our purposes, we are interested in transverse spin states.  If we
choose a frame in which the incoming polarized particle moves along
the $+z$ axis such that $\vec p_\bot = \vec 0_\bot$, then the spinors
(\ref{spinors1}) become eigenstates of the helicity operator.  We can
form transverse spinors as done in \cite{Kovchegov:2012ga} by taking
linear combinations of (\ref{spinors1}) to obtain the projection
along, say, the $x$ axis:
\begin{eqnarray}
\label{spinors2}
U_\chi &= \frac{1}{\sqrt 2} (U_{+z} + \chi U_{-z}) \\ \nonumber
V_\chi &= \frac{1}{\sqrt 2} (V_{+z} - \chi V_{-z}) ,
\end{eqnarray}
where $\chi = \pm 1$ is the spin eigenvalue along the $x$ axis.
Again, for a $\vec p_\bot = \vec 0_\bot$ incoming particle, these
spinors reflect a spin projection transverse to the beam axis; they
are simultaneous eigenstates of the Dirac operator as well as the
Pauli-Lubanski vector $W_1$, where $W_\mu \equiv -\frac{1}{2}
\epsilon_{\mu \nu \rho \sigma} S^{\nu \rho} p^\sigma$.

Using \eqref{CPT1} in \eqref{spinors2} gives the somewhat different
$\mathrm{C/P/T}$ identities satisfied by the transverse spinors:
\begin{eqnarray}
\label{CPT2}
\mathrm{C}: \hspace{1cm} &-i \gamma^2 V_\chi^* (p) = U_{-\chi} (p) \\ \nonumber
\mathrm{CPT}: \hspace{1cm} & U_\chi (p) = - \chi \gamma^5 V_\chi (p) .
\end{eqnarray}
Employing and generalizing  \eqref{CPT2} allows us to write a
complete set of identities for any transverse spinor matrix element:
\begin{eqnarray}
\label{CPT3}
\mathrm{C:} \hspace{1cm}
\vbar{\chi'}(k) \gamma^{\mu_1} \cdots \gamma^{\mu_n} V_{\chi} (p)  
 &=& \left[ \vbar{\chi}(p) \gamma^{\mu_n} \cdots \gamma^{\mu_1} V_{\chi'} (k) \right]^* \\ \nonumber 
 &=& (-1)^{n-1} \ubar{-\chi}(p) \gamma^{\mu_n} \cdots \gamma^{\mu_1} U_{-\chi'}(k) \\ \nonumber
 &=& (-1)^{n-1} \left[ \ubar{-\chi'}(k) \gamma^{\mu_1} \cdots \gamma^{\mu_n} U_{-\chi}(p) \right]^*
\\ \nonumber \\ \label{CPT4}
\mathrm{C:} \hspace{1cm}
\ubar{\chi'}(k) \gamma^{\mu_1} \cdots \gamma^{\mu_n} V_{\chi}(p) 
 &=& \left[ \vbar{\chi}(p) \gamma^{\mu_n} \cdots \gamma^{\mu_1} U_{\chi'} (k) \right]^* \\ \nonumber
 &=& (-1)^{n-1} \ubar{-\chi}(p) \gamma^{\mu_n} \cdots \gamma^{\mu_1} V_{-\chi'}(k) \\ \nonumber
 &=& (-1)^{n-1} \left[ \vbar{-\chi'}(k) \gamma^{\mu_1} \cdots \gamma^{\mu_n} U_{-\chi}(p) \right]^*
\\ \nonumber \\ \label{CPT5}
\mathrm{CPT:} \hspace{1cm}
\vbar{\chi'}(k) \gamma^{\mu_1} \cdots \gamma^{\mu_n} U_\chi (p) 
 &=& \chi \chi' \left[ \vbar{-\chi'}(k) \gamma^{\mu_1} \cdots \gamma^{\mu_n} U_{-\chi}(p) \right]^* \\ \nonumber
\ubar{\chi'}(k) \gamma^{\mu_1} \cdots \gamma^{\mu_n} U_\chi (p)
 &=& \chi \chi' \left[ \ubar{-\chi'}(k) \gamma^{\mu_1} \cdots \gamma^{\mu_n} U_{-\chi}(p) \right]^* .
\end{eqnarray}

These identities allow us to explicitly determine  rigid
constraints on the form of any transverse spinor product.  In
particular, consider the parameterizations of both classes of spinor
products:
\begin{eqnarray}
\label{spinors3}
\vbar{\chi'}(k) \gamma^{\mu_1} \cdots \gamma^{\mu_n} U_\chi(p) \, &\equiv& \,
 \delta_{\chi \chi'} [a(k,p) + \chi a'(k,p)] + \delta_{\chi, -\chi'} [b(k,p) + \chi b'(k,p)] \\ \nonumber
\ubar{\chi'}(k) \gamma^{\mu_1} \cdots \gamma^{\mu_n} U_\chi (p) \, &\equiv& \,
 \delta_{\chi \chi'} [c(k,p) + \chi c'(k,p)] + \delta_{\chi, -\chi'} [d(k,p) + \chi d'(k,p)] ;
\end{eqnarray}
applying \eqref{CPT5}, one readily concludes that $\mathrm{C/P/T}$
constraints imply that:
\begin{itemize}
 \item $a$, $b'$, $c$, and $d'$ are real-valued.
 \item $a'$, $b$, $c'$, and $d$ are pure imaginary.
\end{itemize}

Furthermore, this implies that if we multiply any two of these spinor
matrix elements and sum over one of the spins ($\chi'$), e.g.,
\begin{eqnarray}
\label{spinors4}
& \sum_{\chi'} &  \,  [\vbar{\chi'}(k) \gamma^{\mu_1} \cdots \gamma^{\mu_n} U_\chi(p)] \;
  [\ubar{\chi'}(k) \gamma^{\mu_1} \cdots \gamma^{\mu_n} U_\chi (p)]^* = \\ \nonumber
  &=& \underbrace{[a c^* + a' (c')^* + b d^* + b' (d')^*]}_{\mathrm{real}} \, + \,
      \chi \, \underbrace{[ a (c')^* + a' c^* +  b (d')^* + b' d^* ]}_{\mathrm{imaginary}},
\end{eqnarray}
we find that the product of two transverse matrix elements naturally
partitions into a spin-even, real contribution, and a spin-odd,
imaginary contribution.

Thus in particular, the spin-dependent part of any product of two
transverse matrix elements (say $S_1(\chi)$ and $S_2^*(\chi)$) is
always pure imaginary:
\begin{equation}
\label{spinors5}
S_1 (\chi) S_2^* (\chi) - S_1 (-\chi) S_2^* (-\chi) = - [ S_1^* (\chi) S_2 (\chi) - S_1^* (-\chi) S_2 (-\chi) ].
\end{equation}


\subsection{Relation of the Asymmetry to the Complex Phase}
\label{sec:compphase}

The single transverse spin asymmetry $A_N$ is an observable defined as
\begin{equation}
\label{AN1}
A_N \equiv \frac{d\sigma^\uparrow (\vec{q}_\bot) - d\sigma^\downarrow (\vec{q}_\bot)} {2 \, d\sigma_{unp}} =
 \frac{d\sigma^\uparrow (\tvec{q}) - d\sigma^\uparrow(-\tvec{q})}{2 \, d\sigma_{unp}}
\end{equation}
where $d\sigma(\tvec{q})$ stands for the invariant cross section,
e.g. $\frac{d\sigma}{d^2 q \, dy}$, for the production of a particular
tagged particle with transverse momentum $\tvec{q}$ coming from
scattering on a target with transverse spin $\uparrow , \downarrow$.
Also $d\sigma_{unp}$ represents the normal unpolarized cross section
$\frac{1}{2} (d\sigma^\uparrow + d\sigma^\downarrow)$.  From
\eqref{AN1}, we see that the asymmetry $A_N$ is proportional to the
difference between the amplitude-squared $|\mathcal{A}|^2$ for $\chi =
+1$ and $\chi = -1$.  We denote this spin-difference amplitude squared
as $\Delta |\mathcal{A}|^2$:
\begin{equation}
\label{AN2}
A_N \propto |\mathcal{A}|^2 (\chi = +1) - |\mathcal{A}|^2 (\chi = -1) \equiv \Delta |\mathcal{A}|^2.
\end{equation}

Now let us identify the types of diagrams from which $A_N$ can arise.
Suppose there is a contribution from the square of an amplitude
$\mathcal{A}(\chi) = F \, S(\chi)$ consisting of a spinor product
$S(\chi)$ which depends on the transverse spin eigenvalue $\chi$ and a
factor $F$ coming from the rest of the diagram.  Then the contribution
of the square of the amplitude $\mathcal{A}$ to the asymmetry would be
\begin{eqnarray}
\Delta |\mathcal{A}|^2 &\equiv& |\mathcal{A}|^2 (+1) - |\mathcal{A}|^2 (-1) \\ \nonumber
 &=& |F^2| \, \left[ |S(+1)|^2 - |S(-1)|^2 \right].
\end{eqnarray}
But the constraints on the spinors due to $\mathrm{C/P/T}$
\eqref{spinors5} imply, for $S_1 = S_2 = S$, that
\begin{eqnarray}
\left[ |S(+1)|^2 - |S(-1)|^2 \right] = 0.
\end{eqnarray}
This is easy to understand mathematically: the spin-dependent part
must be pure imaginary, but any amplitude-squared is explicitly real.
Hence, any amplitude squared is independent of $\chi$ and cannot
generate the asymmetry; $A_N$ can only be generated by the quantum
interference between two different diagrams.

Thus at lowest order in perturbation theory, the asymmetry could be
generated by the overlap between an $\mathcal{O}(\alpha_s)$ one-loop
virtual correction and the Born-level amplitude.  So let us consider a
similar exercise to determine the contribution to $\Delta
|\mathcal{A}|^2$ from this $\mathcal{O}(\alpha_s)$ correction (as
compared to the Born-level amplitude squared).  Let us write the
tree-level amplitude $\mathcal{A}_{(0)}$ and the one-loop amplitude
$\mathcal{A}_{(1)}$ as
\begin{eqnarray}
\label{ImPart1}
\mathcal{A}_{(1)}(\chi) &\equiv& F_1 \int d^4 k \frac{S_1(k,\chi)}{D_1(k)} \\ \nonumber
\mathcal{A}_{(0)}(\chi) &\equiv& F_2 \, S_2(\chi)
\end{eqnarray}
where the factor $S_1$ includes all momentum and spin-dependent
numerators, and the factor $D_1$ contains all the propagator
denominators. At $\mathcal{O}(\alpha_s)$ (as compared to the
Born-level amplitude squared) the spin-difference contribution is
\begin{eqnarray}
\label{ImPart2}
\Delta|\mathcal{A}|^2 
 &=& \mathcal{A}_{(1)}(+1) \mathcal{A}_{(0)}^*(+1) + 
  \mathcal{A}_{(1)}^*(+1) \mathcal{A}_{(0)}(+1) \, - \, (\chi \rightarrow - \chi) \\ \nonumber
 &=& F_1 F_2^* \int d^4 k \frac{S_1(k,+1) S_2^*(+1)}{D_1(k)} + 
  F_1^* F_2 \int d^4 k \frac{S_1^*(k,+1) S_2(+1)}{D_1^*(k)} - (\chi \rightarrow -\chi) \\ \nonumber
 &=& F_1 F_2^* \int d^4 k \frac{S_1(k,+1) S_2^*(+1) - S_1(k,-1) S_2^*(-1)}{D_1(k)} 
  + \mathrm{c.c.}.
\end{eqnarray}
But from the $\mathrm{C/P/T}$ constraints \eqref{spinors5}, we see
that the numerator of \eqref{ImPart2} is pure imaginary; thus
\begin{eqnarray}
\label{ImPart3}
\Delta|\mathcal{A}|^2 &=& \int d^4 k \left[ \frac{F_1 F_2^*}{D_1(k)} - \mathrm{c.c.} \right] \,
 \left[S_1(k,+1) S_2^*(+1) - S_1(k,-1) S_2^*(-1)\right] \\ \nonumber
&=& 2 i \int d^4 k \, \mathrm{Im} \left[ \frac{F_1 F_2^*}{D_1(k)} \right] \,
 \left[S_1(k,+1) S_2^*(+1) - S_1(k,-1) S_2^*(-1)\right] .
\end{eqnarray}

Thus we conclude that the spin-dependent part which contributes to the
asymmetry comes only from the imaginary part of the remainder of the 
interference term, aside from the spinor matrix elements.  This 
is easy to understand mathematically: if the spin-dependent part of
the spinor matrix elements is pure imaginary, then it must multiply
the imaginary part of the remainder of the interference term to generate a
real contribution to the asymmetry.  This imaginary part of the rest
of the amplitude interference term gives the complex phase that is
required by $\mathrm{C/P/T}$ to generate the asymmetry $A_N$; it is
not simply the imaginary part of any one diagram, but rather a
relative phase between the tree-level and one-loop amplitudes.  If
there is no relative phase present in the pre-factors,
e.g. $\mathrm{Im}(F_1 F_2^*)=0$, then the imaginary part comes from
the denominator of the loop integral $D_1(k)$.  In that case, taking
the imaginary part corresponds to putting an intermediate virtual
state on energy-shell in light-front perturbation theory.  The
imaginary part generated this way was discussed in
\cite{Brodsky:2002cx} and \cite{Brodsky:2002rv} as a source of the
Sivers-type asymmetry.


\section{Model Calculations with Feynman Diagrams}
\label{ModelFeynman}

Following \cite{Brodsky:2002cx} and \cite{Brodsky:2002rv}, we employ a
toy model of a point-like proton.  To model the parton distribution
function, we introduce a Yukawa-type coupling between the proton
field, the quark field, and a scalar ``diquark'' field.  This
corresponds to an interaction term in the Lagrangian of the form $+G (
\overline{\psi}_p \psi_q + \overline{\psi}_q \psi_p ) \phi_{q
  \overline{q}}$.  The scalar diquark field couples to gluons by the
rules of scalar QED, using the same covariant derivative $i D_\mu =
i \partial_\mu + g A_\mu^a T^a$ as the fermions.  We use $g$ to
represent the QCD coupling of the quark and diquark and $e_f$ to
represent the electromagnetic charge of the quark of flavor $f$.

In the next sections we will explicitly calculate the spin-difference amplitudes at
$\mathcal{O}(\alpha_s)$ defined in \eqref{ImPart2} for deep inelastic
scattering and for the Drell-Yan process.  We will demonstrate in
detail the emergence of the predicted minus sign difference between
the Sivers asymmetries in the two processes.  Throughout the
calculation, we will work with the light-cone coordinates and
corresponding metric
\begin{align}
\label{lcmetric}
p^\pm & \equiv p^{(0)} \pm p^{(3)} 
\\ \nonumber
p^\mu q_\mu & = \frac{1}{2} p^+ q^- + \frac{1}{2} p^- q^+ - \tvec{p} \cdot \tvec{q}
\\ \nonumber
\{\gamma^+ , \gamma^- \} & =  4.
\end{align}
Additionally, we will drop the quark mass $m$ wherever it occurs, but
we will keep the proton mass $M$ and the mass $\lambda$ of the scalar
diquark field.  When it is necessary to approximate the kinematics, we
will work in the limit 
\begin{align}\label{approx}
  s, Q^2 , q_T^2 \gg M^2 , \lambda^2, r_T^2, k_T^2.
\end{align}


\subsection{Semi-Inclusive Deep Inelastic Scattering}
\label{SIDISF}

\begin{figure}
\centering
\includegraphics[width=12cm]{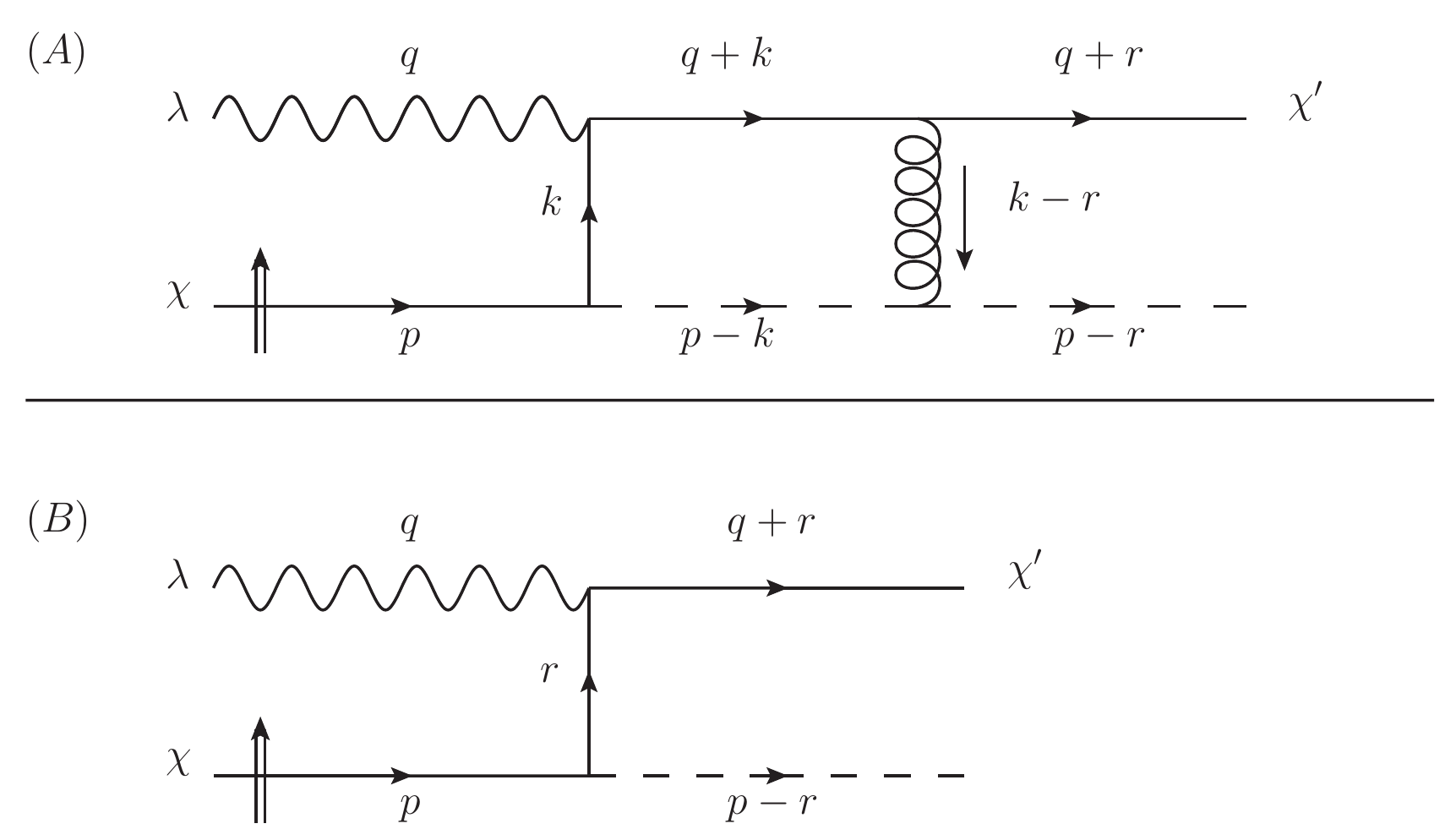}
\caption{Diagrams for the $\gamma^* + p^\uparrow \rightarrow q + X$
  SIDIS amplitude at one-loop order (A) and tree-level (B). The
  incoming solid line denotes the transversely polarized proton, which
  splits into a quark (outgoing solid line) and a diquark (dashed
  line).}
\label{figDIS}
\end{figure}

For deep inelastic scattering, we consider the scattering of a virtual
photon with virtuality $q^2 \equiv - Q^2$ on a transversely polarized
proton with transverse spin eigenvalue $\chi$.  At lowest order, this
process produces a quark and diquark, as shown in Fig. \ref{figDIS}.

To begin, let us establish the kinematics.  Following
\cite{Brodsky:2002cx}, we work in the Drell-Yan-West frame which is
collinear to the proton $(\tvec{p}=\tvec{0})$ and boosted such that
$q^+ =0$ exactly.  In this frame, then, the photon's virtuality comes
from its transverse components: $Q^2 = \tvec{q}^2$.  We define the
longitudinal momentum fraction exchanged in the t-channel as $\Delta
\equiv r^+/ p^+$.  Then momentum conservation and the on-shell
conditions for the proton, quark, and diquark fix $r^-$ and $q^-$:
\begin{eqnarray}
\label{MDISkin}
r^- &=& p^- - (p-r)^- = \frac{M^2}{p^+} - \frac{\tvec{r}^2 + \lambda^2}{(1-\Delta)p^+} \\ \nonumber
q^- &=& (q+r)^- - r^- = \frac{(\tvec{q}+\tvec{r})^2}{\Delta p^+} - r^- 
\approx \frac{Q^2 + 2 \tvec{q} \cdot \tvec{r}}{\Delta p^+} + \mathcal{O}\left(\frac{\bot^2}{p^+}\right).
\end{eqnarray}
These kinematics can be summarized as
\begin{align}
\label{MDISkin2}
p^\mu &= \left( p^+ \, , \, \frac{M^2}{p^+} \, , \, \tvec{0} \right)
\\ \nonumber q^\mu &= \left( 0 \, , \,
  \frac{(\tvec{q}+\tvec{r})^2}{\Delta p^+} - \frac{M^2}{p^+} +
  \frac{\tvec{r}^2 + \lambda^2} {(1-\Delta)p^+} \, , \, \tvec{q}
\right) \\ \nonumber r^\mu &= \left( \Delta p^+ \, , \,
  \frac{M^2}{p^+} - \frac{\tvec{r}^2 + \lambda^2}{(1-\Delta)p^+} \, ,
  \, \tvec{r} \right).
\end{align}
%

With these kinematics, we can write down the one-loop amplitude shown
in Fig.~\ref{figDIS}~(A) as
\begin{eqnarray}
\label{MA1DIS}
\mathcal{A}_1^{DIS} &=& -i g^2 e_f G C_f \int \frac{d^4 k}{(2\pi)^4} \times \\ \nonumber
&\times& \frac{\ubar{\chi'}(q+r) (2 \slashed{p} - \slashed{k} - \slashed{r})(\slashed{q} + \slashed{k}) 
  \slashed{\epsilon}_\lambda \slashed{k} U_\chi (p)} {[k^2 + i\epsilon] \, [(q+k)^2 + i\epsilon] \, [(k-r)^2 + i\epsilon] \, 
  [(p-k)^2 - \lambda^2 + i\epsilon]}
\\ \nonumber
&=&
\frac{i g^2 e_f G C_F}{2 (2\pi)^4 (p^+)^3} \int \frac{dx \, dk^- \, d^2k}{x^2 (x-\Delta) (1-x)} \times 
\\ \nonumber &\times& 
\frac{\ubar{\chi'}(q+r) (2\slashed{p}-\slashed{k}-\slashed{r})(\slashed{q}+\slashed{k})\slashed{\epsilon}_\lambda \slashed{k} U_\chi(p)}{ \left[ k^- - \frac{\tvec{k}^2 - i\epsilon}{x p^+} \right] \left[k^- + q^- - 
    \frac{ (\tvec{q}+\tvec{k})^2-i\epsilon}{x p^+} \right] \left[k^- - r^- - \frac{(\tvec{k}-\tvec{r})^2-i
      \epsilon}{(x-\Delta)p^+} \right] \left[k^- - p^- + \frac{\tvec{k}^2 + \lambda^2 - i\epsilon}{(1-x)p^+} 
  \right]} ,
\end{eqnarray}
where the longitudinal momentum fraction in the loop is $k^+ \equiv x
p^+$ and $C_F = (N_c^2 -1)/2 N_c$ is the Casimir operator in the
fundamental representation.  Similarly, the tree-level amplitude shown
in Fig. \ref{figDIS} (B) is
\begin{align}
\label{MA0DIS}
\mathcal{A}_0^{DIS} = \frac{e_f G}{r^2} \ubar{\chi'}(q+r)
\slashed{\epsilon}_\lambda \slashed{r} U_\chi(p) .
\end{align}

The lowest-order contribution to the spin-difference amplitude squared
comes from the overlap of these diagrams and, in particular, the
imaginary part of the denominators (cf. \eq{ImPart3}):
\begin{eqnarray}
\label{MADIS}
\Delta \left| \mathcal{A}_{DIS} \right|^2 &=& 2i 
\left[ \frac{g^2 e_f^2 G^2 C_F}{2(2\pi)^4 r^2 (p^+)^3} \right]
\int \frac{dx \, d^2k}{x^2 (x-\Delta) (1-x)} \, \mathrm{Im} 
\left\{ \int dk^- \frac{i} {\left[k^- - \frac{\tvec{k}^2-i\epsilon}{x p^+} \right] } 
\right. \\ \nonumber &\times& \left.
\frac{1} 
{
 \left[k^- + q^- - \frac{(\tvec{q}+\tvec{k})^2-i\epsilon}{x p^+} \right] 
 \left[k^- - r^- - \frac{(\tvec{k}-\tvec{r})^2 - i\epsilon}{(x-\Delta)p^+} \right] 
 \left[k^- - p^- + \frac{\tvec{k}^2 + \lambda^2 - i\epsilon}{(1-x)p^+} \right]
}
\right\} \\ \nonumber &\times& \sum_{\chi',\lambda} \left[
\ubar{\chi}(p) \slashed{r} \slashed{\epsilon}^*_\lambda U_{\chi'}(q+r) \ubar{\chi'}(q+r) (2 \slashed{p} - 
\slashed{k} - \slashed{r}) (\slashed{q} + \slashed{k}) \slashed{\epsilon}_\lambda \slashed{k} U_{\chi}(p) 
- (\chi \rightarrow - \chi) \right]
\end{eqnarray}
where we sum over the spin of the outgoing quark and use
\begin{align}
  \label{eq:polsum}
  \sum_\lambda \epsilon^{* \, \mu}_\lambda (q) \, \epsilon^\nu_\lambda
  (q) \rightarrow - g^{\mu\nu}. 
\end{align}
We will provide a justification for summing over photon polarizations
in the two paragraphs following \eq{eq:ANDISDY}. Performing these sums
and simplifying the expression gives
\begin{eqnarray}
\label{MADIS2}
\Delta \left| \mathcal{A}_{DIS} \right|^2 &=& \frac{2i g^2 e_f^2 G^2 C_F}{(2\pi)^4 r^2 (p^+)^3}
\int \frac{dx \, d^2k}{x^2 (x-\Delta)(1-x)} \, \mathcal{I} \,
\\ \nonumber &\times&
 \bigg[ \ubar{\chi}(p) \slashed{r} (\slashed{q} + \slashed{k}) (2 \slashed{p} -
\slashed{k} - \slashed{r}) (\slashed{q} + \slashed{r}) \slashed{k} U_\chi(p) \, - \,
(\chi \rightarrow -\chi) \bigg]
\end{eqnarray}
where the imaginary part that is essential for generating the
asymmetry comes from the expression
\begin{align}
\label{MIDIS}
\mathcal{I} \equiv \mathrm{Im} \! \left\{ \!  \int \! \! \frac{i \;
    dk^-}{ \left[k^- - \frac{\tvec{k}^2-i\epsilon}{x p^+} \right] \!
    \!  \left[k^- + q^- - \frac{(\tvec{q}+\tvec{k})^2-i\epsilon}{x
        p^+} \right] \!  \!  \left[k^- - r^- -
      \frac{(\tvec{k}-\tvec{r})^2-i\epsilon}{(x-\Delta)p^+} \right] \!
    \!  \left[k^- - p^- +
      \frac{\tvec{k}^2+\lambda^2-i\epsilon}{(1-x)p^+} \right] } \! 
\right\} \! .
\end{align}

Notice that the numerator in \eqref{MADIS2} containing the Dirac matrix
element is $k^-$-dependent, such that the $d k^-$ integration
contained in $\mathcal{I}$ from \eq{MIDIS} applies to it
too. Superficially the numerator of \eqref{MADIS2} could
scale as $(k^-)^3$ at large $k^-$, which would endanger convergence;
however, it actually only scales as $(k^-)^2$ since $(\gamma^+)^2 =
0$.  Thus the $k^-$ integral scales at most as $dk^-/(k^-)^{2}$, which
converges and allows us to close the contour in either the upper or
the lower half-plane.

In addition, we will demonstrate below that in the kinematic limit at
hand given by \eq{approx} the leading contribution to the Dirac matrix
element in \eq{MADIS2} is, in fact, $k^-$-independent. We will,
therefore, proceed under the assumption that this is the case and that
all the $k^-$ dependence in \eqref{MADIS2} is contained in the
integrand of \eq{MIDIS}, evaluating the integration in \eqref{MIDIS}
separately.

The imaginary part in \eqref{MIDIS} comes from the denominators, which
corresponds to putting two of the loop propagators on shell: one
occurs from the residue of the $dk^-$ integral and the other occurs by
taking the imaginary part.  However, there are strong kinematic constraints
that restrict which combinations of propagators can go on-shell
simultaneously.  We are working in the limit of massless quarks, and
$1 \leftrightarrow 2$ processes for on-shell massless particles are
forbidden by four-momentum conservation; cuts corresponding to such
processes will explicitly be impossible to put on shell.  Other cuts
correspond to spontaneous proton decay; proton stability against decay
through various channels must be imposed by hand, resulting in
kinematic constraints on the masses of the proton and the scalar.

\eq{MIDIS} is evaluated in Appendix \ref{A}. The result reads
\begin{align}
\label{MIDIS2}
\mathcal{I} = \frac{2\pi^2\Delta^2p^+}{Q^2} \frac{\delta \left[ x -
    \left(1+ 2 \frac{\tvec{q} \cdot ( \tvec{k}-\tvec{r})}{Q^2} \right)
    \Delta \right]}{ \left[ p^- -
    \frac{\tvec{k}^2+\lambda^2}{(1-x)p^+} - \frac{\tvec{k}^2}{xp^+}
  \right] \left[ p^- - r^- - \frac{\tvec{k}^2+\lambda^2}{(1-x)p^+} +
    \frac{(\tvec{k}-\tvec{r})^2}{(\Delta-x)p^+} \right]}.
\end{align}
Substituting this expression back into \eqref{MADIS2} and integrating
over the delta function which sets $x \approx \Delta$ gives
\begin{align}
\label{MADIS3}
\Delta \left|\mathcal{A}_{DIS}\right|^2 & = -i g^2 e_f^2 G^2 C_F
\left(\frac{\Delta(1-\Delta)}{Q^2 (\tvec{r}^2 +a^2)} \right) \int
\frac{d^2k}{(2\pi)^2} \times \\ \nonumber & \times \,
\frac{\ubar{\chi}(p) \slashed{r} (\slashed{q}+\slashed{k}) (2
  \slashed{p}-\slashed{k}-\slashed{r}) (\slashed{q}+\slashed{r})
  \slashed{k} U_\chi (p) - (\chi \rightarrow
  -\chi)}{(\tvec{k}-\tvec{r})^2 (\tvec{k}^2 + a^2)},
\end{align}
where the mass parameter $a^2$ that regulates the infrared divergence
\begin{equation}
\label{Ma2}
a^2 \equiv \Delta \left( \lambda^2 - (1-\Delta)M^2 \right) > 0
\end{equation}
is ensured to be positive definite by the proton stability conditions
\eqref{Mproton1} and \eqref{Mproton2}, and we have used
\begin{equation}
\label{Mr2}
r^2 = \Delta \left( M^2 - \frac{\tvec{r}^2 + \lambda^2}{1-\Delta} \right) - \tvec{r}^2 = 
 - \frac{\tvec{r}^2 + a^2}{1-\Delta}.
\end{equation}
Note that making this cut has fixed the loop momentum $k^\mu$ to be
\begin{equation}
\label{MkDIS}
k^\mu = \left(\Delta p^+ \, , \, \frac{M^2}{p^+} - \frac{\tvec{k}^2+\lambda^2}{(1-\Delta)p^+} \, , \, 
 \tvec{k} \right).
\end{equation}

Next we need to evaluate the numerator of \eqref{MADIS3} by computing
the difference between the matrix elements:
\begin{eqnarray}
\label{NDIS1}
N_{DIS} = \ubar{\chi}(p) \slashed{r} (\slashed{q} + \slashed{k}) (2 \slashed{p} - \slashed{k} - \slashed{r}) (\slashed{q} + \slashed{r}) \slashed{k} U_\chi(p) - (\chi \rightarrow -\chi).
\end{eqnarray}
The momenta involved in this spinor product obey the scale hierarchy
\begin{align}
\label{Nscale}
\overbrace{p^+ , r^+, k^+, q^-, |\tvec{q}|}^{\mathcal{O}(Q)} \, \gg \,
\overbrace{|\tvec{r}|, |\tvec{k}|, M, \lambda}^{\mathcal{O}(\bot)} \,
\gg \, \overbrace{p^- , r^-, k^-}^{\mathcal{O}\left( \bot^2/Q \right)}
,
\end{align}
with the dominant power-counting of the spin-dependent part of the
Dirac matrix element being $\mathcal{O} (Q^4 \bot^2)$. Evaluation of
$N_{DIS}$ in \eq{NDIS1} in the kinematics of \eq{Nscale} is somewhat
involved: after some algebra one can show that there are three classes
of Dirac structures that give a contribution of the leading order,
$\mathcal{O} (Q^4 \bot^2)$; all three involve taking the
$\mathcal{O}(Q)$ momenta from the middle three gamma matrices:
\begin{align}
\label{NDIS2}
N_{DIS} &= \frac{1}{8} \left[ (2p^+ - k^+ - r^+)(q^-)^2 \right] \,
\ubar{\chi}(p) \, \slashed{r} \gamma^+ \gamma^- \gamma^+ \slashed{k} \, U_\chi(p) - 
 (\chi \rightarrow - \chi)
\\ \nonumber &=
\left[(1-\Delta) p^+ (q^-)^2 \right] \,
\ubar{\chi}(p) \, \slashed{r} \gamma^+ \slashed{k} \, U_\chi(p) -  (\chi \rightarrow - \chi).
\end{align}
The three variations consist of taking $\gamma^-$ for both
$\slashed{r}$ and $\slashed{k}$, taking $\gamma^-$ for one and
$\gamma_\bot$ for the other, or taking $\gamma_\bot$ for both.

In the first case, if we take $\gamma^-$ for both $\slashed{r}$ and
$\slashed{k}$, we obtain
\begin{eqnarray}
\label{NDIS3}
\ubar{\chi}(p) \, \slashed{r} \gamma^+ \slashed{k} \, U_\chi(p) &\rightarrow&
\frac{1}{4} \Delta^2 (p^+)^2 \, \ubar{\chi}(p) \, \gamma^- \gamma^+ \gamma^- \, U_\chi(p)
\\ \nonumber &=&
\Delta^2 (p^+)^2 \, \ubar{\chi}(p) \, \gamma^- \, U_\chi(p) ,
\end{eqnarray}
but $\ubar{\chi}(p) \, \gamma^- \, U_\chi(p) = \frac{2M^2}{p^+}$ is
spin-independent and cannot generate the asymmetry.
Similarly, if we take $\gamma_\bot$ for both $\slashed{r}$ and
$\slashed{k}$, we obtain
\begin{eqnarray}
\label{NDIS4}
\ubar{\chi}(p) \, \slashed{r} \gamma^+ \slashed{k} \, U_\chi(p) &\rightarrow& 
r_\bot^i k_\bot^j \ubar{\chi}(p) \, \gamma_\bot^i \gamma^+ \gamma_\bot^j \, U_\chi(p),
\end{eqnarray}
but $\ubar{\chi}(p) \, \gamma_\bot^i \gamma^+ \gamma_\bot^j \,
U_\chi(p) = 2p^+ \delta^{ij}$ is also spin-independent and cannot
generate the asymmetry.
However, if we take one each of $\gamma_\bot$ and $\gamma^-$, we
obtain
\begin{eqnarray}
\label{NDIS5}
\ubar{\chi}(p) \, \slashed{r} \gamma^+ \slashed{k} \, U_\chi(p) &\rightarrow& 
-\frac{1}{2} \Delta p^+ \bigg[
r_\bot^i \ubar{\chi}(p) \, \gamma_\bot^i \gamma^+ \gamma^- \, U_\chi(p) +
\\ \nonumber &+&
k_\bot^i \ubar{\chi}(p) \, \gamma^- \gamma^+ \gamma_\bot^i \, U_\chi(p)
\bigg].
\end{eqnarray}
We can further simplify this expression by using the Dirac equation
\begin{equation}
\label{MDirac}
0 = (\slashed{p}-M) U_\chi(p) = \left[\frac{1}{2}p^+ \gamma^- + \frac{M^2}{2p^+} \gamma^+ - M \right]
U_\chi(p)
\end{equation}
to rewrite the action of $\gamma^-$ in terms of $\gamma^+$ and $M$.  Since $(\gamma^+)^2 = 0$, this simplifies \eqref{NDIS5} to
\begin{eqnarray}
\label{NDIS6}
\ubar{\chi}(p) \, \slashed{r} \gamma^+ \slashed{k} \, U_\chi(p) &\rightarrow& 
- M \Delta (k_\bot^i - r_\bot^i) \, \ubar{\chi}(p) \, \gamma^+ \gamma_\bot^i \, U_\chi(p),
\end{eqnarray}
and $\ubar{\chi}(p) \, \gamma^+ \gamma_\bot^i \, U_\chi(p) = 2 i \chi
p^+ \delta^{i2}$, which is spin-dependent and generates the asymmetry.
Altogether this gives
\begin{eqnarray}
\label{NDIS7}
N_{DIS} &=& -2i\chi \Delta (1-\Delta) (p^+)^2 (q^-)^2 M (k_\bot^{(2)} - r_\bot^{(2)}) - (\chi \rightarrow -\chi)
\\ \nonumber &=&
-4 i \left( \frac{1-\Delta}{\Delta} \right) Q^4 M (k_\bot^{(2)} - r_\bot^{(2)}),
\end{eqnarray}
so that the spin-difference matrix element is pure imaginary, as was
proved in \eqref{spinors5}.

Substituting this result back into \eqref{MADIS3} gives (cf. Eq.~(31)
in \cite{Boer:2002ju}\footnote{As noted in Ref. \cite{Burkardt:2003je}, 
there should be an additional overall minus sign in
front of Eq. (21) of Ref. \cite{Brodsky:2002cx},
also in front of Eq. (31) of Ref. \cite{Brodsky:2002rv}
and Eqs. (31,33,36) of Ref. \cite{Boer:2002ju}.
})
\begin{align}
  \nonumber \Delta \left|\mathcal{A}_{DIS}\right|^2 & = -4 g^2 e_f^2
  G^2 C_F \left(\frac{(1-\Delta)^2 Q^2 M}{\tvec{r}^2 +a^2} \right)
  \int \frac{d^2k}{(2\pi)^2} \frac{k_\bot^{(2)} - r_\bot^{(2)}}
  {(\tvec{k}-\tvec{r})^2 \, (\tvec{k}^2 + a^2)} \\ \label{MADIS4} & =
  +\frac{g^2 e_f^2 G^2 C_F}{\pi} (1-\Delta)^2 \frac{Q^2 M
    r_\bot^{(2)}}{\tvec{r}^2 (\tvec{r}^2 + a^2)} \ln
  \left(\frac{\tvec{r}^2 + a^2}{a^2} \right) ,
\end{align}
where the $d^2k$ integral is performed using Feynman parameters obtaining
\begin{align}
  \label{eq:integral2}
  \int \frac{d^2k}{(2\pi)^2} \frac{k_\bot^{(2)} - r_\bot^{(2)}}
  {(\tvec{k}-\tvec{r})^2 \, (\tvec{k}^2 + a^2)} = - \frac{1}{4 \, \pi} \,
  \frac{r_\bot^{(2)}}{\tvec{r}^2} \, \ln \left(\frac{\tvec{r}^2 +
      a^2}{a^2} \right).
\end{align}
\eq{MADIS4} is the final expression for the spin-difference amplitude
squared for deep inelastic scattering.  We would next like to compare
this expression with the result for the Drell-Yan process.

\begin{figure}
\centering
\includegraphics[width=0.7\textwidth]{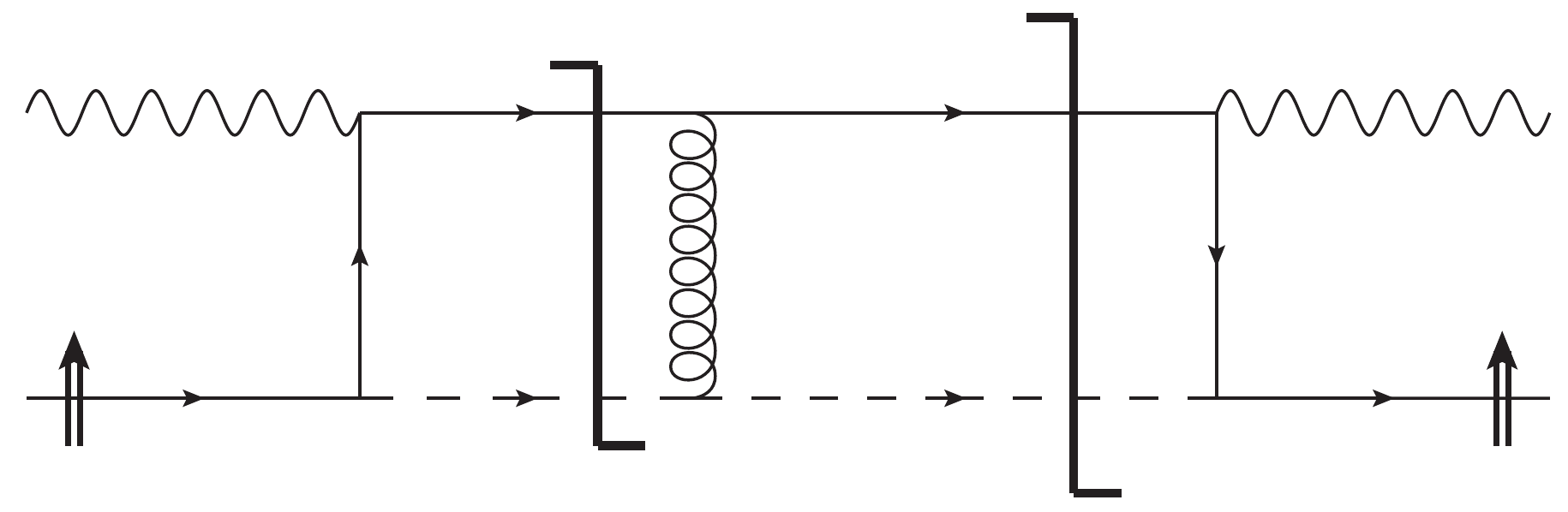}
\caption{Diagrammatic representation of the origin of complex phase
  leading to the single-spin asymmetry in SIDIS. The longer cut
  denotes the final state of the process, while the shorter cut
  demonstrates the origin of the phase needed for the asymmetry.}
\label{DIS-DY2} 
\end{figure}

Before we do that, let us stress once again that the asymmetry in the
SIDIS case arises from the contribution of the diagram in \fig{figDIS}
(A) with the $(q+k)$- and $(p-k)$-lines (corresponding to the lines
labeled \ding{175} and \ding{173} 
in Appendix~\ref{A}) which are put on mass-shell. It is this and
only this contribution that gives the imaginary phase needed for the
asymmetry in SIDIS. This fact becomes more apparent if we diagrammatically
represent putting the $(q+k)$- and $(p-k)$-lines on mass-shell by a
cut, as shown in \fig{DIS-DY2}. In \fig{DIS-DY2} we show the amplitude
squared which we have just calculated, with the longer cut representing the
true final state of the process, and the shorter vertical cut line
representing the imaginary phase generating the asymmetry. The shorter
cut follows the standard Cutkosky rules \cite{Cutkosky:1960sp}, with
the caveat stressed above in Sec.~\ref{sec:compphase} that it should
not be applied to the spinor matrix element; that is, the shorter cut
applies to the denominators of the propagators only, as if we are
evaluating the diagram in a scalar field theory. Using the Cutkosky
rules one can clearly see that this is the only way the shorter cut
line can be placed in the diagram, as all other cuts would lead to
various prohibited $1 \to 2$ or $2 \to 1$ processes, including proton
decay.  Thus \fig{DIS-DY2} demonstrates that the imaginary phase
needed for the single-spin asymmetry arises only in diagrams where it is
possible to place a second cut. We will make use of this result in
the analysis of the Drell-Yan process below.


\subsection{Drell-Yan Process}
\label{sec:DYF}

\begin{figure}[htb]
\centering
\includegraphics[width= 0.8 \textwidth]{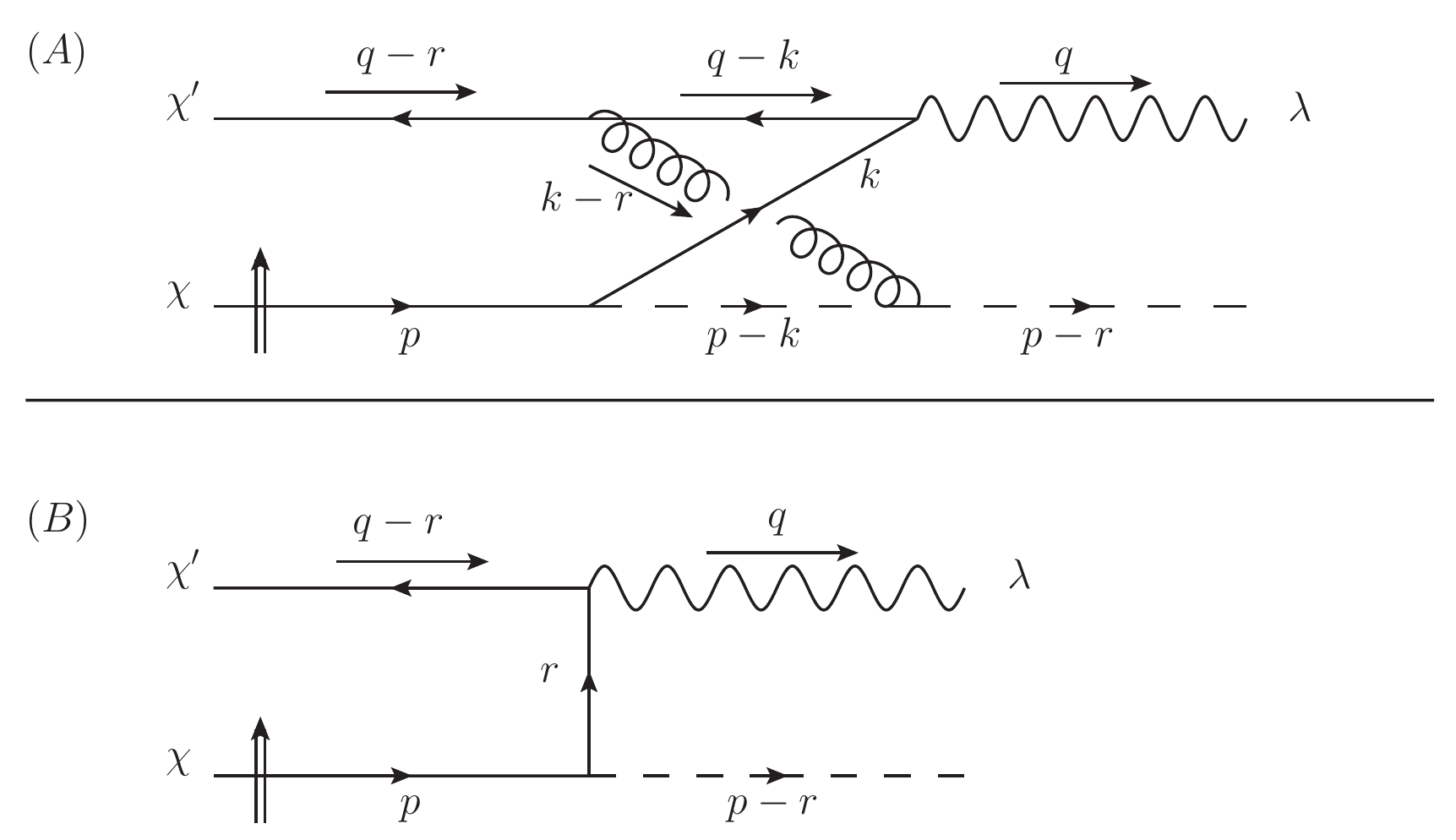}
\caption{Diagrams for the $\overline{q} + p^\uparrow \rightarrow
  \gamma^* + X$ DY amplitude at one-loop order (A) and tree-level
  (B). The incoming proton and anti-quark are denoted by the lower and
  upper solid lines correspondingly, with the outgoing diquark denoted
  by the dashed line.}
\label{figDY}
\end{figure}

We now perform a similar calculation for the Drell-Yan
process in the same model considered above for deep inelastic
scattering.  We will consider the scattering of an antiquark on a
transversely-polarized proton with transverse spin eigenvalue $\chi$
that produces a virtual photon, which then decays into a dilepton pair
with invariant mass $q^2=Q^2$.  This process is shown in
Fig. \ref{figDY} at the level of virtual photon production:
$\overline{q} + p^\uparrow \rightarrow \gamma^* + X$.

Following \cite{Brodsky:2002rv}, we work in a generic frame collinear
to the proton ($\tvec{p}=\tvec{0}$).  We define the longitudinal
momentum fraction of the photon to be $\beta \equiv q^+ / p^+$ and the
momentum fraction exchanged in the t-channel to be $\Delta \equiv r^+
/ p^+$.  As before, four-momentum conservation and the on-shell
conditions fix $r^-$ and $q^-$ to be
\begin{eqnarray}
\label{MDYkin}
r^- &=& p^- - (p-r)^- = \frac{M^2}{p^+} - \frac{\tvec{r}^2 + \lambda^2}{(1-\Delta)p^+} \\ \nonumber
q^- &=& (q-r)^- + r^- = \frac{(\tvec{q}-\tvec{r})^2}{(\beta-\Delta)p^+} + r^- \approx
 \frac{\tvec{q}^2 - 2 \tvec{q}\cdot\tvec{r}}{(\beta-\Delta)p^+} + \mathcal{O}\left(\frac{\bot^2}{p^+}
\right).
\end{eqnarray}
In this frame, the virtual photon's large invariant mass $Q^2$ comes
in part from its transverse components and in part from its longitudinal
components:
\begin{eqnarray}
\label{MQ2DY}
Q^2 \equiv q^2 = \beta p^+ q^- - \tvec{q}^2 \approx \left(\frac{\Delta}{\beta-\Delta}\right) \tvec{q}^2 +
\mathcal{O}\left(\frac{\bot}{Q}\right).
\end{eqnarray}
This allows us to approximate $q^-$ as
\begin{eqnarray}
\label{Mq-DY}
q^- \approx \frac{Q^2}{\Delta p^+} + \mathcal{O}\left(\frac{Q \bot}{p^+}\right),
\end{eqnarray}
which agrees with the corresponding expression \eqref{MDISkin} for DIS
to leading order in $Q^2$. The kinematics can be summarized as
\begin{samepage}
\begin{align}
\label{MDYkin2}
p^\mu &= \left( p^+ \, , \, \frac{M^2}{p^+} \, , \, \tvec{0} \right)
\\ \nonumber q^\mu &= \left( \beta p^+ \, , \,
  \frac{(\tvec{q}-\tvec{r})^2}{(\beta-\Delta)p^+} + \frac{M^2}{p^+} -
  \frac{\tvec{r}^2+\lambda^2}{(1-\Delta)p^+} \, , \, \tvec{q} \right)
\\ \nonumber r^\mu &= \left( \Delta p^+ \, , \, \frac{M^2}{p^+} -
  \frac{\tvec{r}^2 + \lambda^2}{(1-\Delta)p^+} \, , \, \tvec{r}
\right).
\end{align}
\end{samepage}
Notice that, in this frame, the on-shell conditions for the antiquark,
scalar, and dilepton pair imply that $(\beta-\Delta) > 0 , \beta > 0 ,
$ and $(1-\Delta)>0$.  Additionally, to leading order, the positivity
constraint on $q^-$ \eqref{Mq-DY} implies that $\Delta > 0$, and we
can choose 
our frame such that $\beta < 1$, although this is not strictly
necessary.  Altogether, this gives a hierarchy of the fixed scales to
be $0 < \Delta < \beta < 1$.

With these kinematics, we can evaluate the one-loop amplitude shown in
Fig. \ref{figDY} (A) as
\begin{eqnarray}
\label{MA1DY}
\mathcal{A}_1^{DY} &=& \frac{i g^2 e_f G C_F}{(2\pi)^4} \int d^4{k} \frac{ \vbar{\chi'}(q-r) (2\slashed{p} 
- \slashed{k} - \slashed{r})(\slashed{k}-\slashed{q})\slashed{\epsilon}^*_\lambda \slashed{k} U_\chi(p)}
{\left[k^2 + i\epsilon \right] \left[(k-q)^2 + i\epsilon \right] \left[(k-r)^2 + i \epsilon \right]
\left[(p-k)^2 - \lambda^2 + i\epsilon \right]}
\\ \nonumber &=&
\frac{-i g^2 e_f G C_F}{2(2\pi)^4 (p^+)^3} \int \frac{dx \, dk^- \, d^2k}{x(x-\beta)(x-\Delta)(1-x)} \times
\\ \nonumber &\times&
\frac{\vbar{\chi'}(q-r) (2\slashed{p} - \slashed{k} - \slashed{r}) (\slashed{k} - \slashed{q})
\slashed{\epsilon}_\lambda^* \slashed{k} U_\chi(p)}{ \left[ k^- - \frac{\tvec{k}^2-i\epsilon}{xp^+} \right]
\left[k^- - q^- - \frac{(\tvec{k}-\tvec{q})^2-i\epsilon}{(x-\beta)p^+}\right]\left[k^- - r^- - \frac{(\tvec{k}-\tvec{r})^2-i\epsilon}{(x-\Delta)p^+}\right]\left[k^- - p^- + \frac{\tvec{k}^2 + \lambda^2 - i\epsilon}{(1-x)p^+}\right]},
\end{eqnarray}
where $x \equiv k^+ / p^+$ is the longitudinal momentum fraction in
the loop.  Similarly, the tree-level amplitude shown in
Fig. \ref{figDY} (B) is
\begin{align}
\label{MA0DY}
\mathcal{A}_0^{DY} = - \frac{e_f G}{r^2} \vbar{\chi'}(q-r)
\slashed{\epsilon}_\lambda^* \slashed{r} U_\chi(p).
\end{align}
This allows us to calculate the spin-difference amplitude squared
following \eqref{ImPart3} as
\begin{eqnarray}
\nonumber
\Delta |\mathcal{A}_{DY}|^2 &=& 2i \left[\frac{g^2 e_f^2 G^2 C_F}{2(2\pi)^4(p^+)^3 r^2}\right] \int \frac{dx 
\, d^2k} {x(x-\beta)(x-\Delta)(1-x)} \mathrm{Im}\left\{\int dk^- \frac{i}{\left[k^- - 
\frac{\tvec{k}^2-i\epsilon}{xp^+}\right]} \right.
\\ \label{MADY} &\times& \left.
\frac{1}{\left[k^- - q^- - \frac{(\tvec{k}-\tvec{q})^2-i\epsilon}{(x-\beta)p^+} \right] \left[k^- - r^-
- \frac{(\tvec{k}-\tvec{r})^2-i\epsilon}{(x-\Delta)p^+} \right] \left[k^- - p^- + \frac{\tvec{k}^2 +
\lambda^2 - i\epsilon}{(1-x)p^+} \right]} \right\}
\\ \nonumber &\times&
\sum_{\chi' , \lambda} \left[ \ubar{\chi}(p) \, \slashed{r} \slashed{\epsilon}_\lambda V_{\chi'}(q-r) \vbar{\chi'}(q-r) (2\slashed{p} - \slashed{k} - \slashed{r}) (\slashed{k} - \slashed{q}) \slashed{\epsilon}_\lambda^*
\slashed{k} \, U_\chi(p) - (\chi \rightarrow -\chi) \right]
\end{eqnarray}
where we sum over the spin of the incoming antiquark and use
\eq{eq:polsum}. (See the discussion following \eq{eq:ANDISDY} for a
justification of summing over photon polarizations.) Performing these
sums and simplifying the result gives
\begin{align}
\label{MADY2}
\Delta |\mathcal{A}_{DY}|^2 & = \frac{2 i g^2 e_f^2 G^2 C_F}{(2\pi)^4
  r^2 (p^+)^3} \int \frac{dx \, d^2k} {x(x-\beta)(x-\Delta)(1-x)} \,
\mathcal{I} \\ \nonumber & \times \, \left[ \ubar{\chi}(p) \,
  \slashed{r} (\slashed{k} - \slashed{q}) (2\slashed{p} - \slashed{k}
  - \slashed{r}) (\slashed{q}-\slashed{r}) \slashed{k} \, U_\chi(p) -
  (\chi \rightarrow -\chi) \right] ,
\end{align}
where the imaginary part necessary for the asymmetry is generated by
\begin{align}
\label{MIDY}
\mathcal{I} \equiv \mathrm{Im} \! \left\{ \!  \int \! \! \frac{i \;
    dk^-}{ \left[k^- - \frac{\tvec{k}^2-i\epsilon}{xp^+} \right] \! \!
    \left[k^- - q^- -
      \frac{(\tvec{k}-\tvec{q})^2-i\epsilon}{(x-\beta)p^+} \right] \!
    \! \left[k^- - r^- -
      \frac{(\tvec{k}-\tvec{r})^2-i\epsilon}{(x-\Delta)p^+} \right] \!
    \!  \left[k^- - p^- + \frac{\tvec{k}^2 +
        \lambda^2-i\epsilon}{(1-x)p^+} \right] } \!  \right\} \! .
\end{align}
As before, the imaginary part \eqref{MIDY} corresponds to putting two
of the loop propagators on-shell simultaneously: one from performing
the $k^-$ integral and another from taking the imaginary part.  The
propagators that can be simultaneously put on-shell are strongly
constrained by the kinematics and by the requirement of proton
stability. The expression \eqref{MIDY} is evaluated in Appendix
\ref{B} yielding
\begin{align}
\label{MIDY2}
\mathcal{I} = \frac{2\pi^2 \Delta (\beta - \Delta)p^+}{Q^2}
\frac{\delta \left[x - \left(1 + 2 \, \frac{\tvec{q} \cdot (\tvec{k} -
        \tvec{r})} {Q^2} \right) \Delta \right]}{ \left[
    \frac{\tvec{k}^2}{xp^+}-r^- +
    \frac{(\tvec{k}-\tvec{r})^2}{(\Delta-x)p^+} \right] \left[
    \frac{\tvec{k}^2}{xp^+} - p^- + \frac{\tvec{k}^2 +
      \lambda^2}{(1-x)p^+} \right] },
\end{align}
which we can substitute back into \eqref{MADY2}.  Integrating over the
delta function sets $x \approx \Delta$, giving
\begin{align}
\label{MADY3}
\Delta |\mathcal{A}_{DY}|^2 &= -i g^2 e_f^2 G^2 C_F \left(
  \frac{\Delta (1-\Delta)}{Q^2 (\tvec{r}^2 + a^2)} \right) \int
\frac{d^2k}{(2\pi)^2} \\ \nonumber &\times \frac{\ubar{\chi}(p) \,
  \slashed{r} (\slashed{k}-\slashed{q})
  (2\slashed{p}-\slashed{k}-\slashed{r}) (\slashed{q}-\slashed{r})
  \slashed{k} \, U_\chi(p) - (\chi \rightarrow -\chi)} {(\tvec{k} -
  \tvec{r})^2 \, (\tvec{k}^2 + a^2)},
\end{align}
where we have again employed \eqref{Ma2} and \eqref{Mr2}, since we
have established that the proton stability constraint \eqref{Mproton1}
is still valid for the Drell-Yan process.  In performing the
longitudinal integrals, we have fixed the loop momentum $k^\mu$ to be
\begin{equation}
\label{MkDY}
k^\mu = \left(\Delta p^+ \, , \, \frac{\tvec{k}^2}{\Delta p^+} \, , \, \tvec{k} \right) .
\end{equation}

Comparison of \eqref{MADIS3} with \eqref{MADY3} shows that the only
difference between the two processes occurs in the numerators, rather
than in the denominators.  The essential difference in the numerators
is the reversal of the intermediate (anti)quark propagator from
$\slashed{q}+\slashed{k}$ in deep inelastic scattering to $\slashed{k}
- \slashed{q}$ in the Drell-Yan process.  We will return to this point
later in the analysis of the results.


Next we need to evaluate the spin-difference matrix element appearing
in the numerator of \eqref{MADY3}:
\begin{equation}
\label{NDY1}
N_{DY} = \ubar{\chi}(p) \, \slashed{r} (\slashed{k}-\slashed{q}) (2\slashed{p}-\slashed{k}-\slashed{r})
 (\slashed{q}-\slashed{r}) \slashed{k} \, U_\chi(p) - (\chi \rightarrow -\chi).
\end{equation}
The momenta obey the same scale hierarchy \eqref{Nscale} as in deep
inelastic scattering, with the addition of $q^+$ as a scale at
$\mathcal{O}(Q)$ in our frame for Drell-Yan.  The other momenta can
differ from their values in DIS by factors of $\mathcal{O}(1)$, but
the power-counting is the same.  Again, the dominant power-counting of
the matrix element is $\mathcal{O}(Q^4 \bot^2)$, which only arises
from taking
\begin{eqnarray}
\label{NDY2}
(\slashed{k}-\slashed{q}) (2\slashed{p}-\slashed{k}-\slashed{r}) (\slashed{q} - \slashed{r}) &\rightarrow&\
- \frac{1}{8} (q^-)^2 (2p^+ - k^+ - r^+) \, \gamma^+ \gamma^- \gamma^+
\\ \nonumber
&=&
-(1-\Delta) (p^+)(q^-)^2 \gamma^+
\end{eqnarray}
so that
\begin{eqnarray}
\label{NDY3}
N_{DY} = - \left[(1-\Delta) p^+ (q^-)^2 \right] \, \ubar{\chi}(p) \, \slashed{r} \gamma^+ \slashed{k} \,
 U_\chi(p) - (\chi \rightarrow -\chi).
\end{eqnarray}
Comparing \eqref{NDY3} with \eqref{NDIS2}, we see that
\begin{equation}
\label{NDY4}
N_{DY} = - N_{DIS}
\end{equation}
to leading order in $Q$, so we can immediately write the numerator for
Drell-Yan using \eqref{NDIS7} as
\begin{equation}
\label{NDY5}
N_{DY} = +4i \left( \frac{1-\Delta}{\Delta} \right) Q^4 M (k_\bot^{(2)} - r_\bot^{(2)}).
\end{equation}
Substituting this back into \eqref{MADY3} yields the same transverse
momentum integral as in DIS, which we can evaluate using Feynman
parameters to obtain the final answer
\begin{align}
\label{MADY4}
\Delta |\mathcal{A}_{DY}|^2 & = -\frac{g^2 e_f^2 G^2 C_F}{\pi}
(1-\Delta)^2 \frac{Q^2 M r_\bot^{(2)}}{ \tvec{r}^2 (\tvec{r}^2 + a^2)}
\, \ln \left(\frac{\tvec{r}^2 + a^2}{a^2} \right) \\ \nonumber & = -
\Delta |\mathcal{A}_{DIS}|^2.
\end{align}
Thus we conclude that the spin-difference amplitude squared from the
Drell-Yan process is exactly the negative of the that from deep
inelastic scattering, \eqref{MADIS4}.

To obtain the single-spin asymmetry $A_N$ one needs to divide $\Delta
|\mathcal{A}|^2$ for DY and SIDIS by twice the unpolarized amplitude
squared (averaged over the incoming proton polarizations), as follows
from \eq{AN1}. Both in the SIDIS and DY cases the unpolarized
amplitude squared is dominated by the Born-level processes, with the
amplitudes given in Eqs.~\eqref{MA0DIS} and \eqref{MA0DY}
correspondingly. One can easily show that the squares of those
amplitudes, averaged over the proton polarizations, are, in fact,
equal, such that \eq{MADY4} leads to
\cite{Collins:2002kn,Brodsky:2002rv}
\begin{align}
  \label{eq:ANDISDY}
  A_N^{DY} = - A_N^{DIS}.
\end{align}

The sign-reversal has also been derived in \cite{Collins:2002kn} for
the SIDIS and DY Sivers functions
\cite{Sivers:1989cc,Sivers:1990fh,Collins:2002kn,Belitsky:2002sm}. At
the same time in our analysis we have studied the full SIDIS and DY
processes (cf. \cite{Brodsky:2002cx,Brodsky:2002rv}) instead of the
corresponding Sivers functions. However, the conclusion
\eqref{eq:ANDISDY} was reached using \eq{eq:polsum} above: in
particular, note that we have reduced the SIDIS process to $\gamma^* +
p^\uparrow \to q + X$ scattering and have summed over polarizations of
the incoming virtual photon. This is not an exact representation of
the physical SIDIS process, since we have to convolute the hadronic
interaction part of the diagram with a lepton tensor coming from the
electron-photon interactions. Likewise, for the Drell-Yan process we
have replaced the second hadron by an anti-quark, reducing it to the
$q + p^\uparrow \to \gamma^* + X$ scattering. (Replacing the $l^+
l^-$-pair by a time-like photon is also an approximation, true up to
an overall multiplicative factor due to current conservation.) The
resolution for these questions is in the fact that, in the eikonal
kinematics \eqref{approx} considered, the single-spin asymmetries in
the $\gamma^* + p^\uparrow \to q + X$ and $q + p^\uparrow \to \gamma^*
+ X$ processes are proportional to the Sivers functions for SIDIS and
DY correspondingly, as can be shown along the lines of
\cite{Boer:2002ju}.

Consider the quark correlator in a proton
\cite{Boer:2011xd,Boer:2002ju}
\begin{align}
  \label{eq:q_corr}
  \Phi_{ij} (\Delta, {\vec r}_\perp; P, S) \equiv \int \frac{d x^- \,
    d^2 x_\perp}{(2 \, \pi)^3} \, e^{i \, \left(\frac{1}{2} \, x \,
      p^+ \, x^- - {\vec x}_\perp \cdot {\vec r}_\perp \right)} \,
  \langle P, S | {\bar \psi}_j (0) \, {\cal U} \, \psi_i (x^+=0, x^-,
  {\vec x}_\perp) | P, S \rangle ,
\end{align}
where the quark has transverse momentum ${\vec r}_\perp$ and the
longitudinal momentum fraction $\Delta$, the proton spin four-vector
is $S^\mu$, while ${\cal U}$ is the gauge link necessary to make the
object gauge-invariant. The correlation function $\Phi_{ij}$ can be
decomposed as \cite{Boer:1997nt,Boer:2002ju}
\begin{align}
  \label{eq:Phi_dec}
  & \Phi_{ij} (\Delta, {\vec r}_\perp; P, S) = \frac{M}{2 \, p^+} \,
  \bigg[ f_1 (\Delta, {\vec r}_\perp) \, \frac{\slashed{p}}{M} +
  \frac{1}{M^2} \, f_{1 \, T}^\perp (\Delta, {\vec r}_\perp) \,
  \epsilon_{\mu\nu\rho\sigma} \, \gamma^\mu \, p^\nu \, r_\perp^\rho
  \, S_T^\sigma - \frac{1}{M} \, q_{1s} (\Delta, {\vec r}_\perp) \,
  \slashed{p} \, \gamma^5 \notag \\ & - \frac{1}{M} \, h_{1T} (\Delta,
  {\vec r}_\perp) \, i \, \sigma_{\mu\nu} \, \gamma^5 \, S_T^\mu \,
  p^\nu - \frac{1}{M^2} \, h_{1s}^\perp (\Delta, {\vec r}_\perp) \, i
  \, \sigma_{\mu\nu} \, \gamma^5 \, r_\perp^\mu \, p^\nu + h_{1}^\perp
  (\Delta, {\vec r}_\perp) \, \sigma_{\mu\nu} \, \frac{r_\perp^\mu \,
    p^\nu}{M^2} \bigg]_{ij}.
\end{align}
The Sivers function $f_{1 \, T}^\perp (\Delta, {\vec r}_\perp)$ can be
singled out by extracting the spin-dependent part of $\Phi_{ij}
(\gamma^+)_{ji}$, that is \cite{Boer:2002ju}
\begin{align}
  \label{eq:Sivers_ext}
  \Phi_{ij} (\gamma^+)_{ji} \bigg|_{\text{spin dependent}} =
  \frac{2}{M} \, \epsilon^{ij} \, S_T^i \, r_\perp^j \, f_{1 \,
    T}^\perp (\Delta, {\vec r}_\perp).
\end{align}
Comparing \eq{eq:Sivers_ext} to Eqs.~\eqref{NDIS2} and \eqref{NDY3}
above (and comparing the latter two equations to Eq.~(29) in
\cite{Boer:2002ju}) we see that both calculation performed here for
the SIDIS and DY processes single out the corresponding Sivers
functions. In fact our calculation is consistent with that performed
in \cite{Boer:2002ju}, as can be seen by comparing Eqs.~\eqref{MADIS4}
and \eqref{MADY4} to Eq.~(31) in \cite{Boer:2002ju}. We can understand
this consistency from the fact that the summing over the polarization
of the virtual photon after \eq{MADIS} and \eqref{MADY} (see
\eq{eq:polsum}) corresponds to replacing $\gamma^+$ in the left hand
side of \eq{eq:Sivers_ext} by $\gamma_\mu \gamma^+ \gamma^\mu$ which
is just $-2\gamma^+$. Thus the contributions of both
Eqs.~\eqref{MADIS4} and \eqref{MADY4} are proportional to the SIDIS
and DY Sivers functions, calculated in the particular model for the
proton considered here, with, as one can show, identical
proportionality coefficients. (This point is strengthened further by
noticing that the relation in \eq{MADY4} is only valid if one writes
the spin-difference amplitudes in terms of $Q^2$ and $\Delta = x_{F}$,
as is proper for the distribution function like a Sivers function.)
Therefore, the sign reversal in \eq{eq:ANDISDY} is just an explicit
manifestation of the sign reversal between the SIDIS and DY Sivers
functions.

\begin{figure}
\centering
\includegraphics[width=0.7\textwidth]{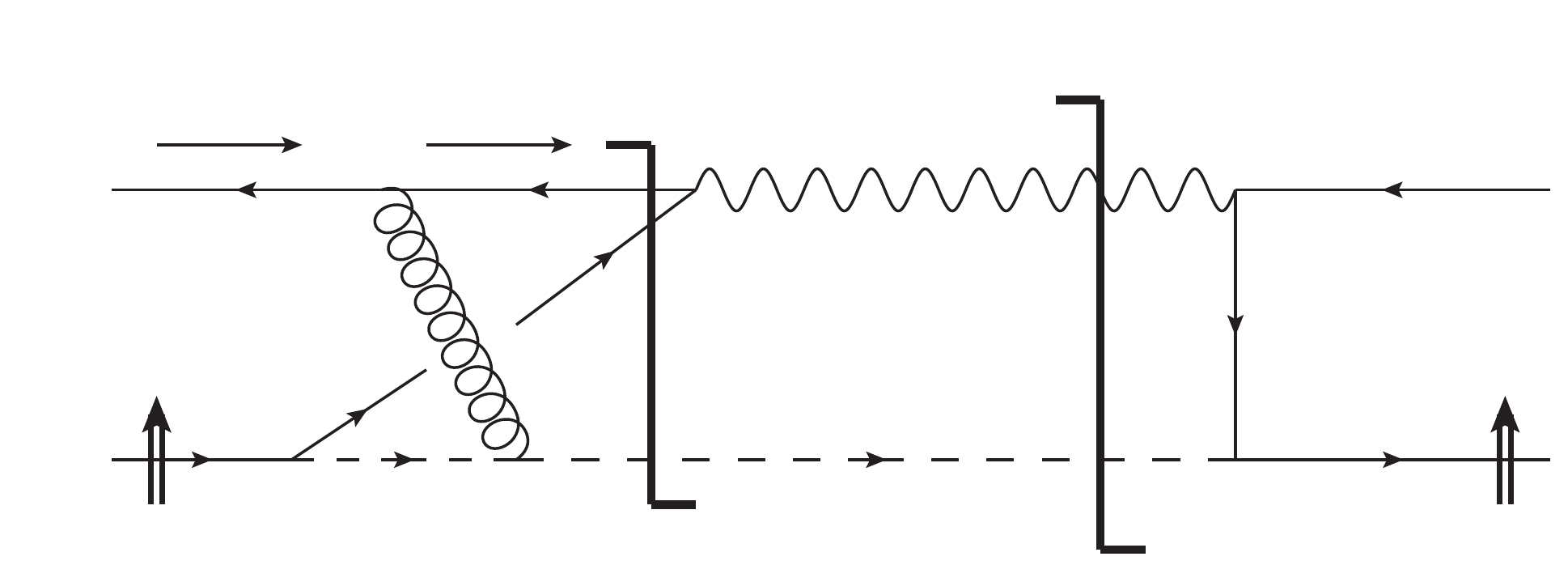}
\caption{Diagrammatic representation of the origin of complex phase
  leading to the single-spin asymmetry in the Drell-Yan process. The
  longer cut denotes the final state of the process, while the shorter
  cut demonstrates the origin of the phase needed for the asymmetry.}
\label{DIS-DY1} 
\end{figure}

It is interesting to investigate the diagrammatic origin of the
sign-flip in Eqs.~\eqref{MADY4} and \eqref{eq:ANDISDY}. To do that we
consider the diagram contributing to the single-spin asymmetry in the
Drell-Yan process shown in \fig{DIS-DY1}. As follows from the
calculation in Appendix~\ref{B}, the asymmetry in the Drell-Yan case
arises due to putting the $(q-k)$- and $k$-lines in \fig{figDY} (A)
(corresponding to lines \ding{172} and \ding{173} in Figs.~\ref{MDYA}
and \ref{MDYB}) on mass-shell: this is illustrated in \fig{DIS-DY1} by
the second (shorter) cut, in analogy to \fig{DIS-DY2}. Comparing
Figures \ref{DIS-DY1} and \ref{DIS-DY2}, we see that the minus sign in
Eqs.~\eqref{MADY4} and \eqref{eq:ANDISDY} arises due to the
replacement of the outgoing eikonal quark in \fig{DIS-DY2} by the
incoming eikonal anti-quark in \fig{DIS-DY1}: this is in complete
analogy with the original Wilson-line time-reversal argument of
Collins \cite{Collins:2002kn} (see also \cite{Belitsky:2002sm}).

However, a closer inspection of Figures \ref{DIS-DY2} and
\ref{DIS-DY1} reveals that the cuts generating the complex phase
appear to be different: in \fig{DIS-DY2} the (shorter) cut crosses the
struck quark and the diquark lines, while in \fig{DIS-DY1} the
(shorter) cut crosses the anti-quark line and the line of the quark in
the proton wave function. While we have already identified the
outgoing quark/incoming anti-quark duality in SIDIS vs. DY as
generating the sign flip, the fact that in the proton's wave function
the diquark is put on mass shell in SIDIS and the quark is put on mass
shell in DY makes one wonder why the absolute magnitudes of the
asymmetries in \eq{eq:ANDISDY} are equal. After all, different cuts may
lead to different contributions to the magnitudes of the asymmetry.
 
\begin{figure}[htb]
\centering
\includegraphics[width=0.3\textwidth]{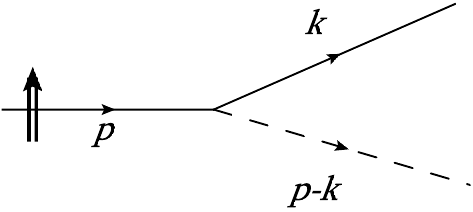}
\caption{Polarized proton splitting into a quark and a diquark, which
  is a part of the diagrams in both Figs.~\protect\ref{DIS-DY2} and
  \protect\ref{DIS-DY1}.}
\label{splitting} 
\end{figure}

Ultimately the origin of \eq{eq:ANDISDY} is in the fact that
spin-asymmetry is a pseudo $T$-odd quantity and the Wilson lines describing
the outgoing quark in SIDIS and the incoming anti-quark in DY are
related by a time-reversal transformation
\cite{Collins:2002kn}. However, in the diagrams at hand the origin of
the equivalence of the shorter cuts in Figs.~\ref{DIS-DY2} and
\ref{DIS-DY1} is as follows. Consider the splitting of a polarized
proton into a quark and a diquark as shown in \fig{splitting}: this
subprocess is common to both diagrams in Figs.~\ref{DIS-DY2} and
\ref{DIS-DY1}. The essential difference between Figs.~\ref{DIS-DY2}
and \ref{DIS-DY1} that we are analyzing is in the fact that in
\fig{DIS-DY2} the diquark is on mass shell, while in \fig{DIS-DY1} the
quark is on mass shell.

Concentrating on the denominators of the quark and diquark propagators
in \fig{splitting} we shall write for the SIDIS case of \fig{DIS-DY2} (quark
is off mass shell, diquark is on mass shell)
\begin{align}
  \label{eq:denom1}
  \frac{1}{k^2} \, \delta \left( (p-k)^2 - \lambda^2 \right) =
  \frac{-1}{p^+ \, ({\vec k}_\perp^2 + a^2)} \, \delta \left( k^- -
    \frac{M^2}{p^+} + \frac{{\vec k}_\perp^2 + \lambda^2}{(1-\Delta)
      \, p^+} \right) \approx \frac{-1}{p^+ \, ({\vec k}_\perp^2 +
    a^2)} \, \delta (k^-),
\end{align}
where we have used Eqs.~\eqref{MDISkin2}, \eqref{Nscale}, and
\eqref{Ma2} along with $x \approx \Delta$, and, in the last step,
neglected all ${\cal O} (\perp^2/Q)$ terms inside the delta-function
since the numerator of the diagram does not depend in the exact value
of $k^-$ as long as it is small.

A similar calculation for the Drell-Yan process from \fig{DIS-DY1}
(quark is on mass shell, diquark is off mass shell in \fig{splitting})
employing Eqs.~\eqref{MDYkin2} and \eqref{Nscale} leads to
\begin{align}
  \label{eq:denom2}
  \frac{1}{(p-k)^2 - \lambda^2} \, \delta \left( k^2 \right) =
  \frac{-1}{p^+ \, ({\vec k}_\perp^2 + a^2)} \, \delta \left( k^- -
    \frac{{\vec k}_\perp^2}{\Delta \, p^+} \right) \approx
  \frac{-1}{p^+ \, ({\vec k}_\perp^2 + a^2)} \, \delta (k^-).
\end{align}
We see that although the two contributions in Eqs.~\eqref{eq:denom1} and
\eqref{eq:denom2} are, in general, different, in the kinematics
\eqref{Nscale} they are apparently equivalent, leading to two
different cuts in Figs.~\ref{DIS-DY2} and \ref{DIS-DY1} giving the
same-magnitude asymmetries.

\begin{figure}[htb]
\centering
\includegraphics[width=0.7\textwidth]{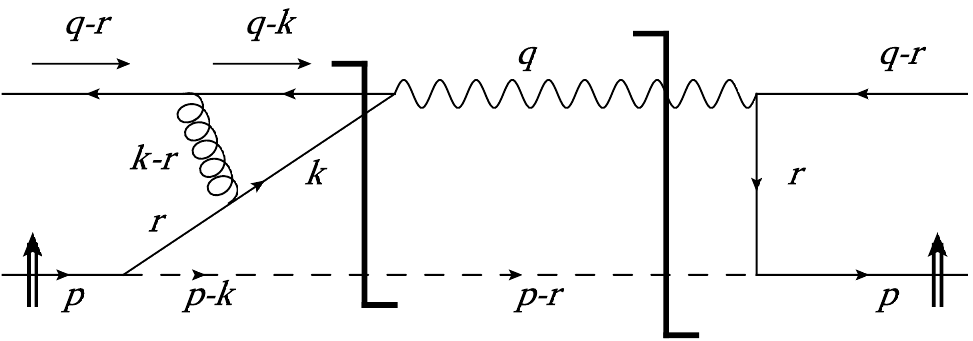}
\caption{The potential contribution to the asymmetry in DY coming from
  the quark-photon vertex correction.}
\label{DY2}
\end{figure}

To complete this Section, let us note that, in the framework of the
model at hand, there is another diagram in the Drell-Yan process which
at first glance contains both the spin-dependence and a complex phase
needed to generate the single-spin asymmetry. The diagram is shown in
\fig{DY2} with its contribution to the single-spin asymmetry denoted
by the double-cut notation of Figs.~\ref{DIS-DY2} and
\ref{DIS-DY1}. The potential contribution to the asymmetry arises due
to a phase generated by the correction to the quark-photon vertex in
\fig{DY2}. Note that an analogous graph cannot give an imaginary part
in the case of SIDIS, since there the virtual photon is space-like.

We will also demonstrate  that the contribution of the diagram in
\fig{DY2} to the single-spin asymmetry is zero. To do this one needs
to evaluate the numerator of this diagram (minus the spin-flip term):
\begin{align}
  \label{eq:num_zero}
  & \sum_\lambda \ubar{\chi}(p) \, \slashed{r}
  \slashed{\epsilon}_\lambda (\slashed{q}-\slashed{r}) \gamma^\mu
  (\slashed{q} - \slashed{k}) \slashed{\epsilon}^*_\lambda \slashed{k}
  \gamma_\mu \slashed{r} \, U_\chi(p) - (\chi \rightarrow -\chi)
  \notag \\ & = - 8 \, k \cdot (q-r) \, \ubar{\chi}(p) \, \slashed{r}
  (\slashed{q} - \slashed{k}) \slashed{r} \, U_\chi(p) - (\chi
  \rightarrow -\chi) \notag \\ & = - 8 \, k \cdot (q-r) \, \left[ 2 \,
    r \cdot (q-k) \, \ubar{\chi}(p) \, \slashed{r} \, U_\chi(p) - r^2
    \, \ubar{\chi}(p) \, (\slashed{q} - \slashed{k}) \, U_\chi(p)
  \right] - (\chi \rightarrow -\chi) = 0.
\end{align}
The zero answer results from the fact that, as can be checked
explicitly, forward Dirac matrix elements of transverse spinors with a
single gamma-matrix, i.e. expressions like $\ubar{\chi}(p) \,
\gamma^\mu \, U_\chi(p)$, are $\chi$-independent. Hence, the diagram
in \fig{DY2} does not contribute to the asymmetry. (In fact, the
second line of \eq{eq:num_zero} is proportional to the square of the
Born term from \fig{figDY} (B): as we show in Sec.~\ref{sec:compphase}
the square of the Born diagram cannot lead to a non-zero single-spin
asymmetry.)

Finally, let us point out that in the calculation of the asymmetries
in both SIDIS and DY, we have neglected diagrams in which the virtual
photon couples to either the proton or the scalar diquark instead of
the (anti)quark. These diagrams are necessary to ensure gauge
invariance, but they are suppressed by powers of $\bot / Q$, which
allowed us to neglect them.


\section{Model Calculations with LFPTH}
\label{ModelLC}

In this section we show that we can re-derive the results obtained in
Section~\ref{ModelFeynman} by using light-front perturbation theory
(LFPTH) \cite{Lepage:1980fj,Brodsky:1997de}. The LFPTH approach and
covariant Feynman diagrams calculations are equivalent; however, we
find it instructive to show the equivalence explicitly.


\subsection{SIDIS}

\begin{figure}[htb]
\centering
\includegraphics[width=0.8 \textwidth]{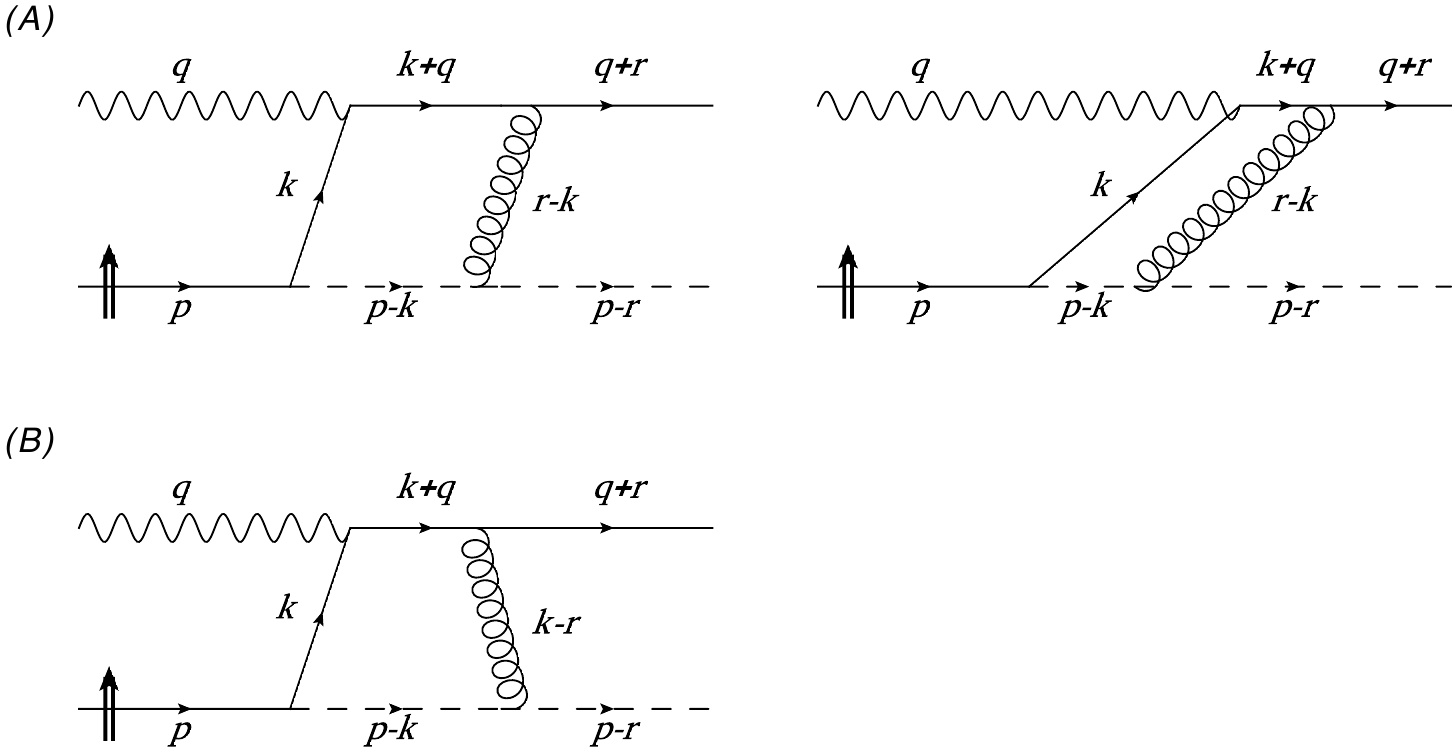}
\caption{The diagrams in LFPTH contributing to the SIDIS amplitude in
  \fig{figDIS} (A). Here labels (A) and (B) denote the diagrams
  corresponding to Cases A and B in Table \ref{MDIStable}.}
\label{lcpt_dis}
\end{figure}

The LFPTH diagrams contributing to the single-spin asymmetry in SIDIS
are shown in \fig{lcpt_dis}. Let us point out from the outset that the
diagrams containing instantaneous terms do not contribute to the
asymmetry in SIDIS and, therefore, are not shown in
\fig{lcpt_dis}. This is clear from the calculation of the numerator in
Sec.~\ref{SIDISF}. The dominant contribution to the numerator comes
from $\gamma^-$ in the quark-gluon vertex, which eliminated the
possibility of the instantaneous gluon exchange contributing. The
factor of $\gamma^+$ arising in the second line of \eq{NDIS2} due to
the photon-quark interactions eliminates the possibility of the
instantaneous quark line exchanges.

The numerators of the amplitudes calculated in the Feynman diagram
approach of Sec.~\ref{ModelFeynman}, consisting of the Dirac matrix
element and vertex factors, are clearly identical to those one would
find using LFPTH. Therefore, we will only study below the energy
denominators of LFPTH diagrams along with the $1/k^+$-factors for
internal lines.

We begin our analysis with Case B from Table \ref{MDIStable} in
Appendix \ref{A}, $x > \Delta$, illustrated in the diagram (B) in
\fig{lcpt_dis}. Concentrating on the light-front energy denominators
and $1/k^+$-factors for internal lines, we write with the help of
\eq{MDISkin2}
\begin{align}
  \label{eq:denoms1}
  & \frac{1}{(k^+)^2 \, (p^+ - k^+) \, (k^+ - r^+)} \, \frac{1}{\left[
      p^- - \frac{\tvec{k}^2}{k^+} - \frac{\tvec{k}^2 + \lambda^2}{p^+
        - k^+} + i \, \epsilon \right] \left[ p^- + q^- -
      \frac{(\tvec{k} + \tvec{q})^2}{k^+} - \frac{\tvec{k}^2 +
        \lambda^2}{p^+ - k^+} + i \, \epsilon \right]} \notag \\ &
  \times \, \frac{1}{\left[ p^- + q^- - \frac{(\tvec{q} +
        \tvec{r})^2}{r^+} - \frac{(\tvec{k} - \tvec{r})^2}{k^+ - r^+}
      - \frac{\tvec{k}^2 + \lambda^2}{p^+ - k^+} + i \, \epsilon
    \right]}.
\end{align}
Employing \eq{MDISkin2} again, which implies the light-cone energy
conservation condition for the diagrams in \fig{lcpt_dis},
\begin{align}
  \label{eq:cons}
  p^- + q^- = \frac{(\tvec{q} + \tvec{r})^2}{r^+} + \frac{\tvec{r}^2 +
    \lambda^2}{p^+ - r^+},
\end{align}
and using the scale hierarchy \eqref{Nscale} to simplify the second
denominator, we recast \eq{eq:denoms1} as
\begin{align}
  \label{eq:denoms2}
  \frac{\Delta \, (1-x) \, (1-\Delta)}{p^+ \, Q^2 \left[ \tvec{k}^2 +
      b^2 \right] \left[ x - \Delta + \frac{2 \, \tvec{q} \cdot (x \,
        \tvec{r} - \Delta \, \tvec{k})}{Q^2} + i \, \epsilon \right]
    \left[ ((1-\Delta) \, \tvec{k} - (1-x) \, \tvec{r})^2 + \lambda^2
      \, (x-\Delta)^2 \right]}
\end{align}
with (cf. \eq{Ma2})
\begin{align}
  \label{eq:b}
  b^2 \equiv x \, \lambda^2 - x \, (1-x) \, M^2 > 0
\end{align}
to impose proton stability (cf. \eqref{Mproton2}). Since the diagram
numerators in the LFPTH and in the above Feynman diagram case are the
same, the argument from Sec.~\ref{sec:compphase} about the need for a
complex phase to generate the spin asymmetry still applies. Therefore
we need to take an imaginary part of \eq{eq:denoms2}, which arises
only from the second denominator, thus putting the intermediate state
involving the $k+q$ and $p-k$ lines in \fig{lcpt_dis} (B) on energy
shell. The imaginary part of \eq{eq:denoms2} is
\begin{align}
  \label{eq:denoms3}
  \frac{- \pi \, \Delta}{p^+ \, Q^2 \left[ \tvec{k}^2 + a^2 \right] \,
    (\tvec{k} - \tvec{r})^2 } \, \delta \left[ x - \left( 1 + \frac{2
        \, \tvec{q} \cdot (\tvec{k} - \tvec{r})}{Q^2} \right) \,
    \Delta \right]
\end{align}
where we also expanded the delta-function prefactor in powers of
$1/Q^2$, which, among other things, put $b=a$.

Comparing \eq{eq:denoms3} to \eq{MIDIS2} (which was also the result of
the diagram evaluation in Case B) or to \eq{MADIS3}, we see that the
denominators of LFPTH give the same structure as the Feynman diagram
calculation.

To study Case A, $x < \Delta$, we analyze the graphs in \fig{lcpt_dis}
(A). We easily observe that the second (right-panel) diagram in
\fig{lcpt_dis} (A) has no imaginary part and can thus be neglected for
our purposes. The difference between the first (left-panel) diagram in
\fig{lcpt_dis} (A) and the graph in \fig{lcpt_dis} (B) is in the third
energy denominator (corresponding to the latest intermediate state),
which, together with the $1/(r^+ - k^+)$ factor, gives
\begin{align}
  \label{eq:denoms4}
  \frac{1}{r^+ - k^+} \, \frac{1}{p^- + q^- - \frac{(\tvec{k} +
      \tvec{q})^2}{k^+} - \frac{(\tvec{r} - \tvec{k})^2}{r^+ - k^+} -
    \frac{\tvec{r}^2 + \lambda^2}{p^+ - r^+} + i \, \epsilon} \notag
  \\ = \frac{- x \, \Delta}{\left[ (\Delta - x) \, \tvec{q} + \Delta
      \, \tvec{k} - x \, \tvec{r} \right]^2} \approx \, \frac{- 1}{(
    \tvec{k} - \tvec{r} )^2}
\end{align}
where, in the last step, we have used the delta-function from
\eq{eq:denoms3} common to the imaginary part of both diagrams along
with the scale hierarchy \eqref{Nscale}. We see that the third
intermediate state gives the same contribution to the imaginary parts
of the diagrams in \fig{lcpt_dis} (A) and (B): hence \eq{eq:denoms3}
is also valid in Case A. 

This completes our demonstration of the equivalence of the LFPTH
calculation for the single-spin asymmetry in SIDIS to the Feynman
diagram calculation.

\subsection{DY}

\begin{figure}[htb]
\centering
\includegraphics[width=0.7 \textwidth]{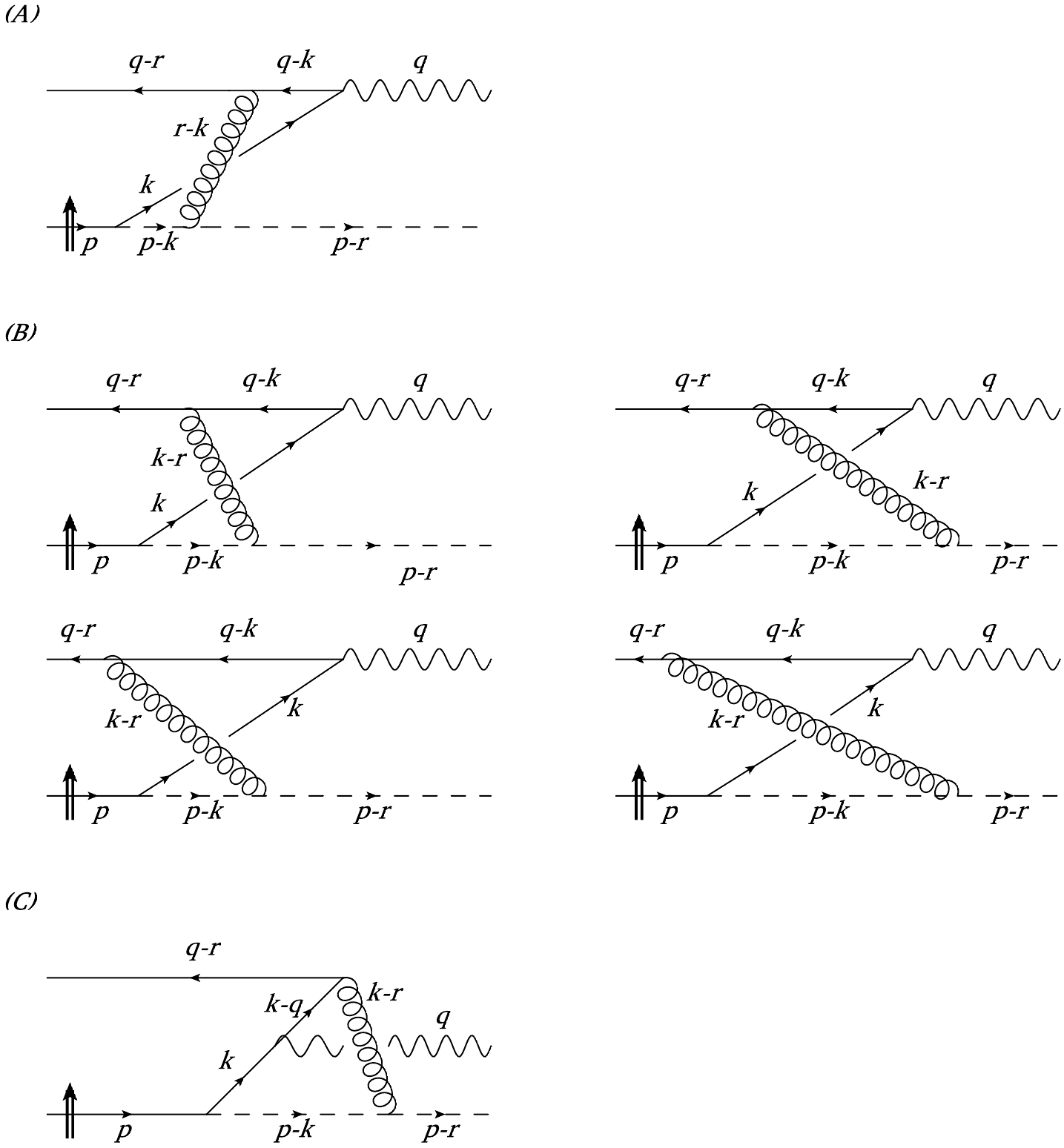}
\caption{The diagrams in LFPTH contributing to the Drell-Yan amplitude
  in \fig{figDY} (A). Here labels (A), (B) and (C) denote the diagrams
  corresponding to Cases A, B and C in Table~\ref{MDYtable} of
  Appendix~\ref{B}. The labels on all the lines indicate the momentum
  flowing to the right.}
\label{lcpt_dy}
\end{figure}

The LFPTH analysis of the Drell-Yan process proceeds along the lines
similar to the SIDIS case. The LFPTH diagrams contributing to the
phase-generating amplitude in \fig{figDY} (A) are shown in
\fig{lcpt_dy} and are labeled (A), (B) and (C) according to the three
non-trivial cases listed in Table~\ref{MDYtable} of Appendix~\ref{B}.

Starting with Case A from Table~\ref{MDYtable}, $0 < x < \Delta$, we
write the contribution of the light-cone energy denominators and
$1/k^+$-type factors for the diagram in \fig{lcpt_dy} (A) as
\begin{align}
  \label{eq:denoms_dy1}
  & \frac{1}{k^+ \, (r^+ - k^+) \, (q^+ - k^+) \, (p^+ - k^+)} \,
  \frac{1}{\left[ p^- - \frac{\tvec{k}^2}{k^+} - \frac{\tvec{k}^2 +
        \lambda^2}{p^+ - k^+} + i \, \epsilon \right] \left[ p^- -
      \frac{\tvec{k}^2}{k^+} - \frac{(\tvec{r} - \tvec{k})^2}{r^+ -
        k^+} - \frac{\tvec{r}^2 + \lambda^2}{p^+ - r^+} + i \,
      \epsilon \right] } \notag \\ & \times \, \frac{1}{\left[ p^- +
      \frac{(\tvec{q} - \tvec{r})^2}{q^+ - r^+} -
      \frac{\tvec{k}^2}{k^+} - \frac{(\tvec{q} - \tvec{k})^2}{q^+ -
        k^+} - \frac{\tvec{r}^2 + \lambda^2}{p^+ - r^+} + i \,
      \epsilon \right]}.
\end{align}
Using the kinematics in \eq{MDYkin2} leading to the light-cone energy
conservation condition
\begin{align}
  \label{eq:e-cons}
  p^- + \frac{(\tvec{q} - \tvec{r})^2}{q^+ - r^+} = q^- +
  \frac{\tvec{r}^2 + \lambda^2}{p^+ - r^+}.
\end{align}
we rewrite the imaginary part of \eq{eq:denoms_dy1} in the following
form
\begin{align}
  \label{eq:denoms5}
  \frac{- \pi \, \Delta}{p^+ \, Q^2 \left[ \tvec{k}^2 + a^2 \right] \,
    (\tvec{k} - \tvec{r})^2 } \, \delta \left[ x - \left( 1 + \frac{2
        \, \tvec{q} \cdot (\tvec{k} - \tvec{r})}{Q^2} \right) \,
    \Delta \right]
\end{align}
in agreement with the factors in \eq{MIDY2} and/or in \eq{MADY3}.

The analysis in Case B, $\Delta < x < \beta$, is slightly more
involved. The denominators of all four graphs in \fig{lcpt_dy} (B)
combine to give
\begin{align}
  \label{eq:denoms-dy2}
  & \left[ \frac{1}{p^- - k^- - (p-k)^- + i \, \epsilon} +
    \frac{1}{(q-r)^- - (q-k)^- - (k-r)^- + i \, \epsilon} \right]
  \notag \\ & \times \, \frac{1}{p^- + (q-r)^- - (q-k)^- - (k-r)^- -
    k^- - (p-k)^- + i \, \epsilon} \notag \\ & \times \, \left[
    \frac{1}{q^- - k^- - (q-k)^- + i \, \epsilon} + \frac{1}{(p-r)^- -
      (k-r)^- - (p-k)^- + i \, \epsilon} \right] \notag \\ & = \,
  \frac{1}{\left[ p^- - k^- - (p-k)^- + i \, \epsilon \right] \left[
      (q-r)^- - (q-k)^- - (k-r)^- + i \, \epsilon \right]} \notag \\
  & \times \, \left\{ \frac{1}{(p-r)^- - (k-r)^- - (p-k)^- + i \,
      \epsilon} + \frac{1}{q^- - (q-k)^- - k^- + i \, \epsilon}
  \right\}.
\end{align}
The first term in the curly brackets of \eq{eq:denoms-dy2}, multiplied
by the other two denominators, contains no imaginary part, since each
energy denominator in that term represents a forbidden $1\to 2$ decay
or a forbidden $2 \to 1$ merger. Only the second term in the curly
brackets of \eq{eq:denoms-dy2}, corresponding to the diagrams on the
left of \fig{lcpt_dy} (B), has an imaginary part. A simple calculation
shows that this imaginary part (after it is multiplied by the
$1/k^+$-type terms) is equal to \eq{eq:denoms5}, thus extending its
validity into the $\Delta < x < \beta$ region in agreement with our
Feynman diagram calculations.

Finally noticing that the diagram in \fig{lcpt_dy} (C) has no imaginary
part, just like in the Feynman diagram case, we complete the
demonstration of the equivalence of the LFPTH calculation for the
single-spin asymmetry in DY to the Feynman diagram calculation.

\section{Conclusions}
\label{sec:conclusions}

In this paper we have calculated the single transverse spin
asymmetries for the $\gamma^* + p^\uparrow \to q + X$ and ${\bar q} +
p^\uparrow \to \gamma^* + X$ processes in the model where the proton
consists of a quark and a scalar diquark. We have shown explicitly
that the SSAs arise from different cuts in the two processes in the
Feynman diagram language (see Figs.~\ref{DIS-DY2} and \ref{DIS-DY1}),
corresponding to putting different energy denominators on energy shell
in the LFPTH formalism. In spite of this difference, in the end of the
calculation we get a simple sign-flip relation between spin
asymmetries in the two processes, \eq{eq:ANDISDY}, in agreement with
the arguments based on time-reversal anti-symmetry of the SSA
\cite{Collins:2002kn}.  The detailed calculation is consistent with
the underlying dynamics of the lensing effect shown in \fig{relsign}.
for QED. The final-state interaction in SIDIS is attractive, whereas
the initial state interaction is repulsive in DY. Note that the Sivers
effect is leading twist in $Q^2$ even though the virtuality of the
exchanged gluon which appears in the lensing effect is small.  This is
consistent with the OPE which is valid for small values of the ratios
$M^2/Q^2$ and $r_{\perp}^2/Q^2$.

The origin of the sign reversal at the diagrammatic level is discussed at
the end of Sec.~\ref{sec:DYF}, following \fig{splitting}. It appears
from this discussion that the sign-flip  between the SSAs in
SIDIS and DY may only hold for the large-$s$ and large-$Q^2$
kinematics considered here and given by \eq{approx}. While this was
not checked explicitly in this work, it appears that the sign-flip
relation \eqref{eq:ANDISDY} may not hold outside of this
approximation, and may thus be destroyed by corrections to the
transverse-momentum distribution (TMD) factorization used in
\cite{Collins:2002kn}. More work is needed to investigate this further.

To summarize, we have confirmed the sign-flip relation
\eqref{eq:ANDISDY} by an explicit diagrammatic calculation in a simple
and robust model, in the process finding subtleties in the
diagrammatic representation of \eqref{eq:ANDISDY} which were not known
before.


\section*{Acknowledgments}

Yu.K. is grateful to Leonard Gamberg, Jianwei Qiu, and Feng Yuan for
discussions which inspired him to revisit the problem.  The research
of S.J.B. was supported in part by the Department of Energy contract
DE-AC02-76SF00515; the research of D.S.H. is supported in part by the
Korea Foundation for International Cooperation of Science \&
Technology (KICOS) and the Basic Science Research Programme through
the National Research Foundation of Korea (2012-0002959); the research
of Yu.K. and M.S. is sponsored in part by the U.S. Department of
Energy under Grant No. DE-SC0004286; the research of I. S. is
supported by Project Basal under Contract No. FB0821, and by the
Fondecyt project 1100287.


\appendix

\renewcommand{\theequation}{A\arabic{equation}}
 \setcounter{equation}{0}
 \section{Integration in the SIDIS Case}
\label{A}


Our goal in this Appendix is to evaluate the expression
\eqref{MIDIS}. There are four poles to the $dk^-$ integral, labeled
below as \ding{172} - \ding{175}.  Depending on the hierarchy of the
longitudinal momentum fractions $x$ and $\Delta$, these poles may be
located either above or below the real $k^-$ axis.  Since the outgoing
quark and scalar are on-shell, we have $(q+r)^+ = \Delta p^+ > 0$ and
$(p-r)^+ = (1-\Delta)p^+ > 0$ so that $0<\Delta<1$.  This allows us to
write four distinct kinematic regimes in which to classify the poles:
$(x < 0 < \Delta < 1)$ , $(0 < x < \Delta < 1)$, $(0 < \Delta < x <
1)$, and $(0 < \Delta < 1 < x)$.  The classification of the four pole
locations as above or below the real $k^-$ axis for each of these
regimes is listed in Table \ref{MDIStable}.

\begin{table}[h]
\centering
\begin{tabular}{|cl|c|c|c|c|}
 \hline &Pole & $x<0$ & $0<x<\Delta<1$ & $0<\Delta<x<1$ & $x>1$ \\ \hline
 \ding{172}& $k^- = \frac{\tvec{k}^2-i\epsilon}{xp^+}$ & above & below & below & below \\
 \ding{173}& $k^- = - q^- + \frac{(\tvec{q}+\tvec{k})^2-i\epsilon}{xp^+}$ & above & below & below & below \\
 \ding{174}& $k^- = r^- + \frac{(\tvec{k}-\tvec{r})^2-i\epsilon}{(x-\Delta)p^+}$ & above & above & below &
  below \\
 \ding{175}& $k^- = p^- - \frac{\tvec{k}^2 + \lambda^2 -i\epsilon}{(1-x)p^+}$ & above & above & above &
  below \\ \hline \hline
 &Contribution: & $0$ & Case A & Case B & $0$ \\ \hline
\end{tabular}
\caption{Table classifying the pole locations of \eqref{MIDIS} as lying either above or below the $\mathrm{Re} \, k^-$ axis.}
\label{MDIStable} 
\end{table}

For $x<0$ or $x>1$, all the poles fall on the same side of the
$\mathrm{Re}\, k^-$ axis, so that we can close the contour in the
other direction and get zero contribution.  The physical region
corresponds to $0<x<1$, and there are two distinct time-orderings of
the diagram, $x<\Delta$ and $x>\Delta$.  We examine these two cases
below.

\begin{figure}[htb]
\centering
\includegraphics[width=0.9\textwidth]{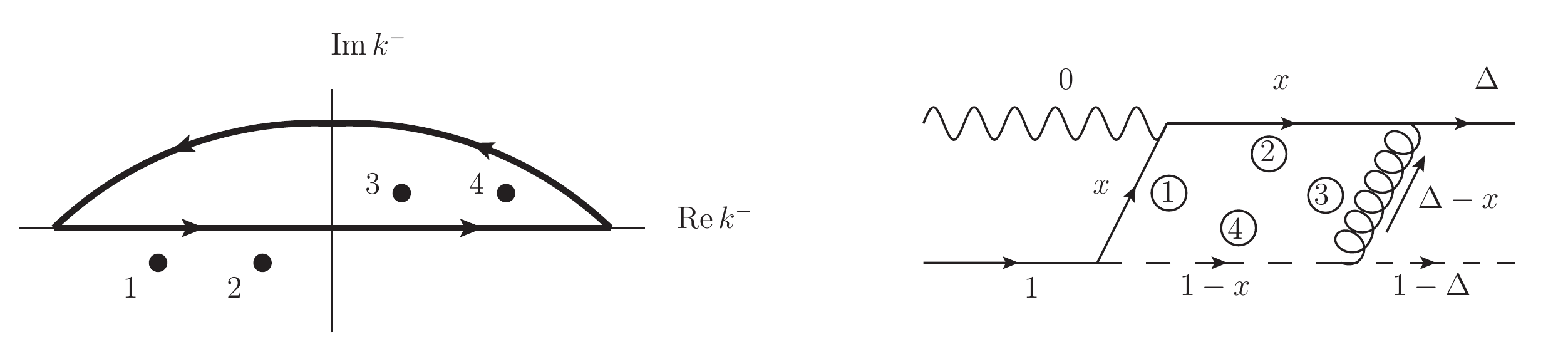}
\caption{Illustration of the poles (left) and
  corresponding time-ordered diagram (right) of \eqref{MIDIS} for the
  kinematic regime Case A: $0<x<\Delta<1$.  We choose to close the
  contour in the upper half-plane, enclosing the poles \ding{174} and
  \ding{175}.  Note that the placement of the poles is only schematic,
  indicating the sign of their imaginary part; the placement on the
  real axis has no significance.}
\label{MDISA} 
\end{figure}

\vspace{0.5cm}

For Case A: $0<x<\Delta<1$, we can close the contour in the upper
half-plane, enclosing the poles \ding{174} and \ding{175}, as in
Fig. \ref{MDISA}.  Let us consider the possible contributions to
\eqref{MIDIS} from the residue and imaginary parts of the various
poles.

\begin{itemize}
 \item \underline{Res[\ding{174}] Im[\ding{173}]: Kinematically Prohibited}
\end{itemize}

This term would yield a contribution of
\begin{eqnarray}
\label{Mcut1}
\mathcal{I} &=& \frac{-2\pi \, \mathrm{Im} \left\{ \frac{1}{r^- + q^- - \frac{(\tvec{k}-\tvec{r})^2}
{(\Delta-x)p^+} -\frac{(\tvec{q}+\tvec{k})^2}{xp^+} +i\epsilon} \right\}
}{\left[r^- - \frac{(\tvec{k}-\tvec{r})^2}{(\Delta-x)p^+} - \frac{\tvec{k}^2}
{xp^+} \right] \left[ r^- - p^- - \frac{(\tvec{k}-\tvec{r})^2}{(\Delta-x)p^+}+\frac{\tvec{k}^2+\lambda^2}
{(1-x)p^+} \right]} 
\\ \nonumber &=&
\frac{+2\pi^2 \, \delta\left[ r^- + q^- - \frac{(\tvec{k}-\tvec{r})^2} {(\Delta-x)p^+} -\frac{(\tvec{q}+
\tvec{k})^2}{xp^+} \right]} {\left[r^- - \frac{(\tvec{k}-\tvec{r})^2}{(\Delta-x)p^+} - \frac{\tvec{k}^2}
{xp^+} \right] \left[ r^- - p^- - \frac{(\tvec{k}-\tvec{r})^2}{(\Delta-x)p^+}+\frac{\tvec{k}^2+\lambda^2}
{(1-x)p^+} \right]}
\\ \nonumber \mathcal{I} &=&
\frac{+2\pi^2 \frac{x\Delta}{\Delta-x} p^+ \, \delta \left[ \left(
\tvec{q}+\frac{\Delta}{\Delta-x}\tvec{k}-\frac{x}{\Delta-x}\tvec{r} \right)^2 \right]}
{\left[r^- - \frac{(\tvec{k}-\tvec{r})^2}{(\Delta-x)p^+} - \frac{\tvec{k}^2}
{xp^+} \right] \left[ r^- - p^- - \frac{(\tvec{k}-\tvec{r})^2}{(\Delta-x)p^+}+\frac{\tvec{k}^2+\lambda^2}
{(1-x)p^+} \right]},
\end{eqnarray}
but the argument of the delta function is positive definite, so it
cannot be satisfied; this cut is kinematically prohibited because it
corresponds to a 2 $\rightarrow$ 1 massless, on-shell process.

\begin{itemize}
 \item \underline{Res[\ding{174}] Im[\ding{172}]: Proton Decay}
\end{itemize}

Similarly, this cut would yield a contribution of
\begin{eqnarray}
\label{Mcut2}
\mathcal{I} &=& 
\frac{ +2\pi^2 \, \delta 
\left[ 
 r^- - \frac{(\tvec{k}-\tvec{r})^2}{(\Delta-x)p^+} - \frac{\tvec{k}^2}{xp^+} 
\right] }{
\left[
 r^- + q^- - \frac{(\tvec{k}-\tvec{r})^2}{(\Delta-x)p^+} - \frac{(\tvec{q}+\tvec{k})^2}{xp^+}
\right] 
\left[
 r^- - p^- - \frac{(\tvec{k}-\tvec{r})^2}{(\Delta-x)p^+} + \frac{\tvec{k}^2 + \lambda^2}{(1-x)p^+}
\right]}
\\ \nonumber \mathcal{I} &\propto&
\delta \bigg[ 
 -x(\Delta-x)\left(\lambda^2-(1-\Delta)M^2\right) - x(\Delta-x)\tvec{r}^2 - x(1-\Delta)(\tvec{k}-\tvec{r})^2
 \\ \nonumber &-& (1-\Delta)(\Delta-x)\tvec{k}^2
\bigg].
\end{eqnarray}
All of the terms inside the delta-function are negative definite except
for the first one, so we can impose the stability of the proton by
requiring that
\begin{equation}
\label{Mproton1}
\lambda^2 - (1-\Delta)M^2 > 0.
\end{equation}

\begin{itemize}
\item \underline{Res[\ding{174}] Im[\ding{175}] + Res[\ding{175}]
    Im[\ding{174}]: Kinematically Prohibited (Cancels)}
\end{itemize}

The term corresponding to Res[\ding{174}] Im[\ding{175}] is
\begin{eqnarray}
\label{Mcut3}
\mathcal{I}_1 &=& 
\frac{ -2\pi \, \mathrm{Im} 
\left\{ 
 \frac{1}{
 r^- - p^- - \frac{(\tvec{k}-\tvec{r})^2-i\epsilon}{(\Delta-x)p^+} + \frac{\tvec{k}^2 + \lambda^2-i\epsilon}
 {(1-x)p^+}
} \right\} }{
\left[
 r^- - \frac{(\tvec{k}-\vec{r})^2}{(\Delta-x)p^+} - \frac{\tvec{k}^2}{xp^+}
\right] 
\left[
 r^- + q^- - \frac{(\tvec{k}-\tvec{r})^2}{(\Delta-x)p^+} - \frac{(\tvec{q}+\tvec{k})^2}{xp^+}
\right]}
\\ \nonumber &=&
\frac{ \mp 2\pi^2 \, \delta 
\left[ 
 r^- - p^- - \frac{(\tvec{k}-\tvec{r})^2}{(\Delta-x)p^+} + \frac{\tvec{k}^2 + \lambda^2}{(1-x)p^+}
\right] }{
\left[
 r^- - \frac{(\tvec{k}-\vec{r})^2}{(\Delta-x)p^+} - \frac{\tvec{k}^2}{xp^+}
\right] 
\left[
 r^- + q^- - \frac{(\tvec{k}-\tvec{r})^2}{(\Delta-x)p^+} - \frac{(\tvec{q}+\tvec{k})^2}{xp^+}
\right]}.
\end{eqnarray}
The sign of the $i\epsilon$ argument of the imaginary part is
ambiguous; this is typically a signature of a false pole.  Whatever
the sign of this term, it is exactly canceled by the Res[\ding{175}]
Im[\ding{174}] term:
\begin{eqnarray}
\label{Mcut4}
\mathcal{I}_2 &=& 
\frac{ -2\pi \, \mathrm{Im} 
\left\{ 
 \frac{1}{
 p^- - \frac{\tvec{k}^2 + \lambda^2 -i\epsilon}{(1-x)p^+} - r^- + \frac{(\tvec{k}-\tvec{r})^2-i\epsilon}
 {(\Delta-x)p^+}
} \right\} }{
\left[
 p^- - \frac{\tvec{k}^2 + \lambda^2}{(1-x)p^+} - \frac{\tvec{k}^2}{xp^+}
\right] 
\left[
 p^- + q^- - \frac{\tvec{k}^2 + \lambda^2}{(1-x)p^+} - \frac{(\tvec{q}+\tvec{k})^2}{xp^+}
\right]}
\\ \nonumber &=&
\frac{ \pm 2\pi^2 \, \delta 
\left[ 
 p^- - \frac{\tvec{k}^2 + \lambda^2}{(1-x)p^+} - r^- + \frac{(\tvec{k}-\tvec{r})^2}{(\Delta-x)p^+}
\right] }{
\left[
 p^- - \frac{\tvec{k}^2 + \lambda^2}{(1-x)p^+} - \frac{\tvec{k}^2}{xp^+}
\right] 
\left[
 p^- + q^- - \frac{\tvec{k}^2 + \lambda^2}{(1-x)p^+} - \frac{(\tvec{q}+\tvec{k})^2}{xp^+}
\right]}.
\end{eqnarray}
Thus this cut, which would correspond to a massless, on-shell $1
\rightarrow 2$ process, is kinematically prohibited.

\begin{itemize}
\item \underline{Res[\ding{175}] Im[\ding{172}]: Proton Decay}
\end{itemize}

This cut corresponds to proton decay through a different channel,
yielding a contribution of
\begin{eqnarray}
\label{Mcut5}
\mathcal{I} &=& 
\frac{ +2\pi^2 \, \delta 
\left[ 
 p^- - \frac{\tvec{k}^2+\lambda^2}{(1-x)p^+} - \frac{\tvec{k}^2}{xp^+}
\right] }{
\left[
 p^- + q^- - \frac{\tvec{k}^2+\lambda^2}{(1-x)p^+} - \frac{(\tvec{q}+\tvec{k})^2}{xp^+}
\right] 
\left[
 p^- - r^- - \frac{\tvec{k}^2 + \lambda^2}{(1-x)p^+} + \frac{(\tvec{k}-\tvec{r})^2}{(\Delta - x)p^+}
\right]}
\\ \nonumber \mathcal{I} &\propto&
\delta \bigg[ 
 -x \left(\lambda^2 - (1-x) M^2 \right) - \tvec{k}^2
\bigg].
\end{eqnarray}
To prevent proton decay through this channel, we need to impose the
slightly different condition
\begin{equation}
\label{Mproton2}
\lambda^2 - (1-x) M^2 > 0
\end{equation}

\begin{itemize}
 \item \underline{Res[\ding{175}] Im[\ding{173}]: Legal Cut}
\end{itemize}

This combination is the only legal cut of the four denominators that
can be put on-shell simultaneously.  This contribution is
\begin{eqnarray}
\label{Mcut6}
\mathcal{I} &=& 
\frac{ +2\pi^2 \, \delta 
\left[ 
 p^- + q^- - \frac{\tvec{k}^2 + \lambda^2}{(1-x)p^+} - \frac{(\tvec{q}+\tvec{k})^2}{xp^+}
\right] }{
\left[
 p^- - \frac{\tvec{k}^2+\lambda^2}{(1-x)p^+} - \frac{\tvec{k}^2}{xp^+}
\right] 
\left[
 p^- - r^- - \frac{\tvec{k}^2+\lambda^2}{(1-x)p^+} + \frac{(\tvec{k}-\tvec{r})^2}{(\Delta-x)p^+}
\right]}.
\end{eqnarray}
Expanding the argument of the delta function and keeping terms of
order $\mathcal{O}\left(\frac{\bot}{Q}\right)$ gives
\begin{align}\label{delta_full}
  \delta \left[ p^- + q^- - \frac{\tvec{k}^2 + \lambda^2}{(1-x)p^+} -
    \frac{(\tvec{q}+\tvec{k})^2}{xp^+} \right] \approx
  \frac{\Delta^2 p^+}{Q^2} \, \delta \left[ x - \left(1+ 2
      \frac{\tvec{q} \cdot ( \tvec{k}-\tvec{r})}{Q^2} \right) \Delta
  \right].
\end{align}
The delta function sets $x \approx \Delta$ to leading order, but the
singularity of the delta function only falls within the kinematic
region of Case A, $0<x<\Delta<1$ if
\begin{equation}
\nonumber
\tvec{q} \cdot (\tvec{k} - \tvec{r}) <0 ,
\end{equation}
which is restricted to only half of the total phase space of the
$d^2k$ integral.  As we will see, Case B complements this integral
with the other half of the phase space.  With this caveat, we can
write a final expression for the imaginary part as
\begin{align}
\label{MIDIS22}
\mathcal{I} = \frac{2\pi^2\Delta^2p^+}{Q^2} \frac{\delta \left[ x -
    \left(1+ 2 \frac{\tvec{q} \cdot ( \tvec{k}-\tvec{r})}{Q^2} \right)
    \Delta \right]}{ \left[ p^- -
    \frac{\tvec{k}^2+\lambda^2}{(1-x)p^+} - \frac{\tvec{k}^2}{xp^+}
  \right] \left[ p^- - r^- - \frac{\tvec{k}^2+\lambda^2}{(1-x)p^+} +
    \frac{(\tvec{k}-\tvec{r})^2}{(\Delta-x)p^+} \right]}.
\end{align}

\vspace{0.5cm}

\begin{figure}[h]
\centering
\includegraphics[width=\textwidth]{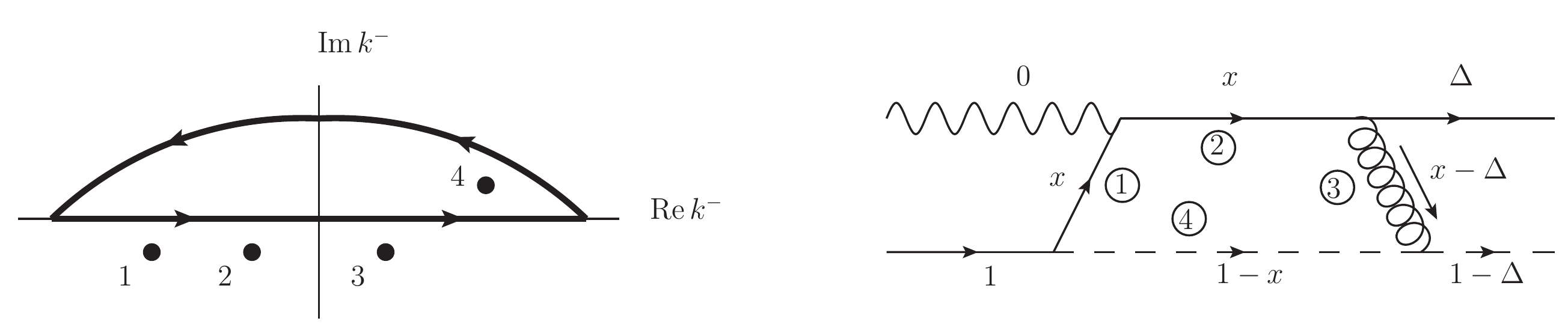}
\caption{Illustration of the poles (left) and corresponding
  time-ordered diagram (right) of \eqref{MIDIS} for the kinematic
  regime Case B: $0<\Delta<x<1$.  We choose to close the contour in
  the upper half-plane, enclosing only the pole \ding{175}.  Note that
  the placement of the poles is only schematic, indicating the sign of
  their imaginary part; the placement on the real axis has no
  significance.}
\label{MDISB} 
\end{figure}

For Case B: $0 < \Delta < x < 1$, we can close the contour in the
upper half-plane, enclosing only the pole \ding{175}, as in
Fig. \ref{MDISB}.  Again, we will consider the various contributions
to \eqref{MIDIS} from the residue and imaginary part of the various
poles.

\begin{itemize}
 \item \underline{Res[\ding{175}] Im[\ding{172}]: Proton Decay}
\end{itemize}

This cut is exactly the same as the corresponding cut \eqref{Mcut5} in
Case A; it is unaffected by changing the sign of $(x-\Delta)$.  Thus
the condition to prohibit proton decay through this channel is the
same: $\lambda^2 - (1-x)M^2 >0$.

\begin{samepage}
\begin{itemize}
 \item \underline{Res[\ding{175}] Im[\ding{174}]: Kinematically Prohibited}
\end{itemize}
In this regime, it is explicitly impossible to perform this cut:
\end{samepage}
\begin{eqnarray}
\label{Mcut7}
\mathcal{I} &=& 
\frac{ +2\pi^2 \, \delta 
\left[ 
 p^- - r^- - \frac{\tvec{k}^2 + \lambda^2}{(1-x)p^+} - \frac{(\tvec{k}-\tvec{r})^2}{(x-\Delta)p^+}
\right] }{
\left[
 p^- - \frac{\tvec{k}^2 + \lambda^2}{(1-x)p^+} - \frac{\tvec{k}^2}{xp^+}
\right] 
\left[
 p^- + q^- - \frac{\tvec{k}^2 + \lambda^2}{(1-x)p^+} - \frac{(\tvec{q}+\tvec{k})^2}{xp^+}
\right]}
\\ \nonumber \mathcal{I} &\propto&
\delta \bigg[ 
 -(x-\Delta)^2 \lambda^2 - \left( (1-x)\tvec{r} + (1-\Delta)\tvec{k} \right)^2
\bigg].
\end{eqnarray}
Since the argument is negative definite, this cut is kinematically prohibited.

\begin{itemize}
 \item \underline{Res[\ding{175}] Im[\ding{173}]: Legal Cut}
\end{itemize}

Again, this is the only combination of propagators that can be put on
shell simultaneously.  The expression is the same as in \eqref{Mcut6}
but with $x>\Delta$.  This means that the delta function
\begin{eqnarray}
  \nonumber
  \delta\left[
    x- \left(1+2 \frac{\tvec{q}\cdot(\tvec{k}-\tvec{r})}{Q^2}  \right) \, \Delta
  \right]
\end{eqnarray}
has its singularity within the kinematic window of Case B, $0 < \Delta
< x < 1$, if
\begin{eqnarray}
\nonumber
\tvec{q}\cdot(\tvec{k}-\tvec{r}) >0.
\end{eqnarray}

Thus Case B gives rise to the same expression \eqref{MIDIS22} at
leading order, but with validity in the complementary region of the
$d^2k$ phase space; the final expression \eqref{MIDIS22} is thus valid
for all $\tvec{k}$. 

The expression \eqref{MIDIS22} is illustrated in \fig{DIS-DY2} in
terms of the extra cut corresponding to putting the quark and the
diquark propagators on mass shell.



\renewcommand{\theequation}{B\arabic{equation}}
 \setcounter{equation}{0}
 \section{Integration in the DY Case}
\label{B}


Here we want to evaluate the expression in \eqref{MIDY}.  In Table
\ref{MDYtable} we classify the four poles \ding{172} - \ding{175} of
this expression as lying either above or below the $\mathrm{Re} \,
k^-$ axis for the five distinct kinematic regimes:
$(x<0<\Delta<\beta<1)$, $(0<x<\Delta<\beta<1)$,
$(0<\Delta<x<\beta<1)$, $(0<\Delta<\beta<x<1)$, and
$(0<\Delta<\beta<1<x)$.  As before all the poles lie to one side of
the real axis unless $0<x<1$, so there are three distinct cases to
evaluate, each of which corresponds to a particular time-ordering of
the diagram.  We consider each of these cases below.
%
\begin{table}[h]
\centering
\begin{tabular}{|cl|c|c|c|c|c|}
 \hline &Pole & $x<0$ & $0<x<\Delta$ & $\Delta<x<\beta$ & $\beta<x<1$ & $x>1$ 
  \\ \hline
 \ding{172}& $k^- = \frac{\tvec{k}^2-i\epsilon}{xp^+}$ & above & below & below & below & below \\
 \ding{173}& $k^- = q^- + \frac{(\tvec{k}-\tvec{q})^2-i\epsilon}{(x-\beta)p^+}$ & above & above & above & 
  below & below \\
 \ding{174}& $k^- = r^- + \frac{(\tvec{k}-\tvec{r})^2-i\epsilon}{(x-\Delta)p^+}$ & above & above & below &
  below & below \\
 \ding{175}& $k^- = p^- - \frac{\tvec{k}^2 + \lambda^2 -i\epsilon}{(1-x)p^+}$ & above & above & above &
  above & below \\ \hline \hline
 &Contribution: & $0$ & Case A & Case B & Case C & $0$ \\ \hline
\end{tabular}
\caption{\label{MDYtable} Table classifying the pole locations of \eqref{MIDY} as lying either above or below the $\mathrm{Re} \, k^-$ axis.}
\end{table}
%

\vspace{0.5cm}

\begin{figure}[htb]
\centering
\includegraphics[width=\textwidth]{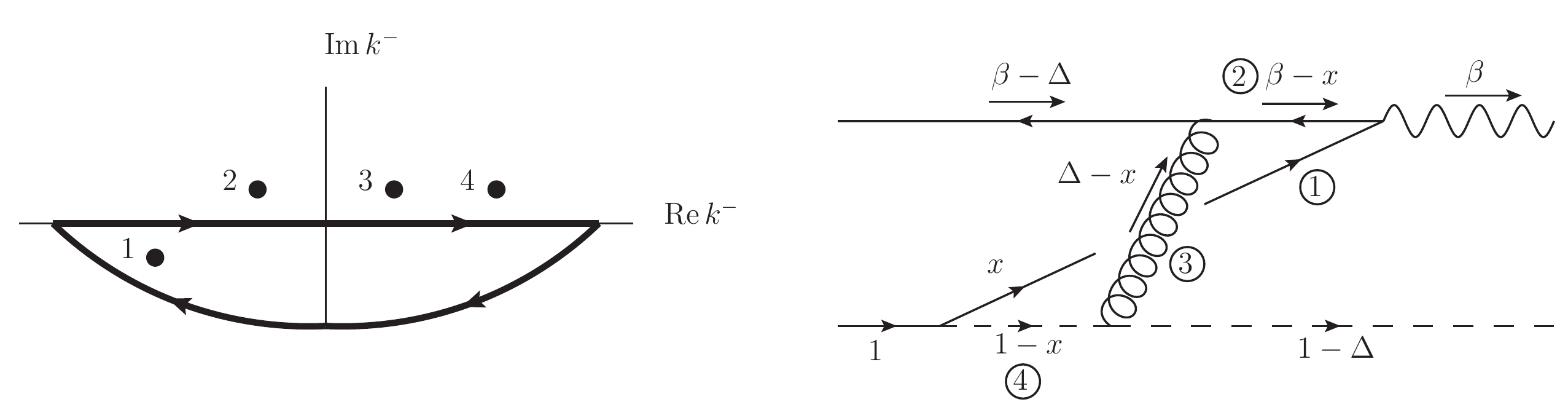}
\caption{\label{MDYA} Illustration of the poles (left) and
  corresponding time-ordered diagram (right) of \eqref{MIDY} for the
  kinematic regime Case A: $0<x<\Delta<\beta<1$.  We choose to close
  the contour in the lower half-plane, enclosing only the pole
  \ding{172}.  Note that the placement of the poles is only schematic,
  indicating the sign of their imaginary part; the placement on the
  real axis has no significance.}
\end{figure}

For Case A: $0<x<\Delta<\beta<1$, we choose to close the contour in
the lower half-plane, enclosing only the pole \ding{172}, as shown in
Fig. \ref{MDYA}.  Let us consider the possible contributions to
\eqref{MIDY} from the residue and imaginary parts of the various
poles.

\begin{itemize}
 \item \underline{ Res[\ding{172}] Im[\ding{175}]: Proton Decay }
\end{itemize}
This contribution would give
\begin{eqnarray}
\label{Mcut8}
\mathcal{I} &=& \frac{+2\pi^2 \delta
 \left[
  \frac{\tvec{k}^2}{xp^+} - p^- + \frac{\tvec{k}^2 + \lambda^2}{(1-x)p^+}
 \right] }{
 \left[
  \frac{\tvec{k}^2}{xp^+} - q^- + \frac{(\tvec{k}-\tvec{q})^2}{(\beta-x)p^+}
 \right]
 \left[
  \frac{\tvec{k}^2}{xp^+} - r^- + \frac{(\tvec{k}-\tvec{r})^2}{(\Delta-x)p^+}
 \right]}
\\ \nonumber &\propto&
 \delta \left[
 x \left(\lambda^2 - (1-x) M^2 \right) + \tvec{k}^2
 \right],
\end{eqnarray}
so we can prohibit proton decay through this channel by requiring that
$\samepage{\lambda^2 - (1-x)M^2 > 0}$.

\begin{itemize}
 \item \underline{ Res[\ding{172}] Im[\ding{174}]: Proton Decay }
\end{itemize}
This contribution would give
\begin{eqnarray}
\label{Mcut9}
\mathcal{I} &=& \frac{+2\pi^2 \delta
 \left[
  \frac{\tvec{k}^2}{xp^+} - r^- + \frac{(\tvec{k}-\tvec{r})^2}{(\Delta-x)p^+}
 \right] }{
 \left[
  \frac{\tvec{k}^2}{xp^+}-q^- + \frac{(\tvec{k}-\tvec{q})^2}{(\beta-x)p^+}
 \right]
 \left[
  \frac{\tvec{k}^2}{xp^+}-p^- + \frac{\tvec{k}^2 + \lambda^2}{(1-x)p^+}
 \right]}
\\ \nonumber &\propto&
 \delta \bigg[
  x(\Delta-x) \left(\lambda^2 - (1-\Delta)M^2 \right) + (1-\Delta)(\Delta-x) \tvec{k}^2 + x(\Delta-x) 
  \tvec{r}^2 + 
\\ \nonumber &+& 
 x(1-\Delta) (\tvec{k}-\tvec{r})^2 
 \bigg].
\end{eqnarray}
All of the momenta are positive definite, so we can prohibit proton
decay through this channel by requiring that $\samepage{\lambda^2 -
  (1-\Delta)M^2 > 0}$.

\begin{itemize}
 \item \underline{ Res[\ding{172}] Im[\ding{173}]: Legal Cut }
\end{itemize}
This corresponds to the only legal cut of the diagram as shown in
Fig. \ref{MDYA}; it is permitted because it corresponds to a $2
\rightarrow 1$ process in which the two massless quarks become a
single ``massive'' time-like photon with ``mass'' $Q$.  Equivalently,
we can recognize that the subsequent leptonic decay of the time-like
virtual photon makes this cut correspond to a massless, on-shell $2
\rightarrow 2$ scattering process, which is allowed.  This cut makes a
contribution of
\begin{eqnarray}
\label{Mcut10}
\mathcal{I} &=& \frac{+2\pi^2 \delta
 \left[
  \frac{\tvec{k}^2}{xp^+} - q^- + \frac{(\tvec{k}-\tvec{q})^2}{(\beta-x)p^+}
 \right] }{
 \left[
  \frac{\tvec{k}^2}{xp^+} - r^- + \frac{(\tvec{k}-\tvec{r})^2}{(\Delta-x)p^+}
 \right]
 \left[
  \frac{\tvec{k}^2}{xp^+} - p^- + \frac{\tvec{k}^2 + \lambda^2}{(1-x)p^+}
 \right]},
\end{eqnarray}
and
\begin{align}
  \nonumber \delta\left[\frac{\tvec{k}^2}{xp^+} - q^- +
    \frac{(\tvec{k}-\tvec{q})^2}{(\beta-x)p^+}\right] \approx
  \frac{\Delta(\beta-\Delta)p^+}{Q^2} \ \delta \left[x - \left(1 + 2
      \, \frac{\tvec{q} \cdot (\tvec{k} - \tvec{r})} {Q^2} \right)
    \Delta \right].
\end{align}
As usual, the $\delta$-function sets $x \approx \Delta$, but the
singularity only falls within the kinematic window of Case A
$(x<\Delta)$ for $\tvec{q} \cdot (\tvec{k}-\tvec{r})<0$.  As with DIS,
this half of the $d^2k$ phase space will be complemented by an equal
contribution for Case B $\Delta < x < \beta$.  Thus, the legal cut
gives
\begin{align}
\label{MIDY22}
\mathcal{I} = \frac{2\pi^2 \Delta (\beta - \Delta)p^+}{Q^2}
\frac{\delta \left[x - \left(1 + 2 \, \frac{\tvec{q} \cdot (\tvec{k} -
        \tvec{r})} {Q^2} \right) \Delta \right]}{ \left[
    \frac{\tvec{k}^2}{xp^+}-r^- +
    \frac{(\tvec{k}-\tvec{r})^2}{(\Delta-x)p^+} \right] \left[
    \frac{\tvec{k}^2}{xp^+} - p^- + \frac{\tvec{k}^2 +
      \lambda^2}{(1-x)p^+} \right] }
\end{align}

\vspace{0.5cm}
%
\begin{figure}
\centering
\includegraphics[width=\textwidth]{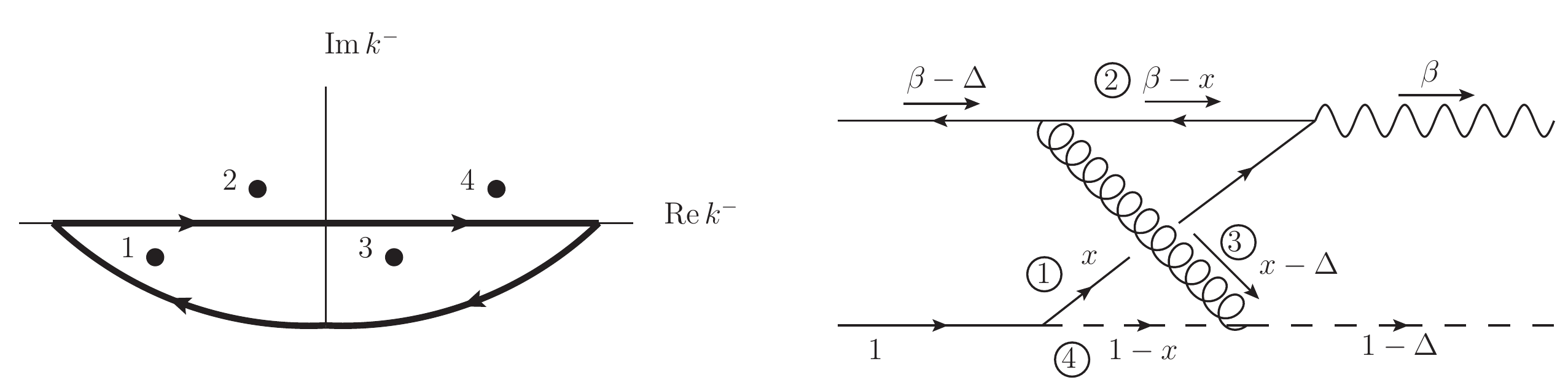}
\caption{\label{MDYB} Illustration of the poles (left) and
  corresponding time-ordered diagram (right) of \eqref{MIDY} for the
  kinematic regime Case B: $0<\Delta<x<\beta<1$.  We choose to close
  the contour in the lower half-plane, enclosing the poles \ding{172}
  and \ding{174}.  Note that the placement of the poles is only
  schematic, indicating the sign of their imaginary part; the
  placement on the real axis has no significance.}
\end{figure}
%
For Case B: $0<\Delta<x<\beta<1$, we close the contour in the lower
half-plane, enclosing the poles \ding{172} and \ding{174}, as shown in
Fig. \ref{MDYB}.  Let us consider the possible contributions to
\eqref{MIDY} from the residue and imaginary parts of the various
poles.
\begin{itemize}
 \item \underline{ Res[\ding{172}] Im[\ding{175}]: Proton Decay }
\end{itemize}
The evaluation of this cut proceeds along exactly the same lines as in
\eqref{Mcut8} of Case A; the proton stability condition is unaffected
by changing the sign of $(\Delta-x)$.

\begin{itemize}
 \item \underline{ Res[\ding{172}] Im[\ding{174}] + Res[\ding{174}] Im[\ding{172}]: Proton Decay (Cancels)}
\end{itemize}
Evaluating Res[\ding{172}] Im[\ding{174}] would give a contribution of
\begin{eqnarray}
\label{Mcut11}
\mathcal{I}_1 &=& \frac{+2\pi \, \mathrm{Im} \left\{ \frac{1}{
  \frac{\tvec{k}^2-i\epsilon}{xp^+} - r^- - \frac{(\tvec{k}-\tvec{r})^2-i\epsilon}{(x-\Delta)p^+}
 } \right\} }{
 \left[
  \frac{\tvec{k}^2}{xp^+}-q^- + \frac{(\tvec{k}-\tvec{q})^2}{(\beta-x)p^+}
 \right]
 \left[
  \frac{\tvec{k}^2}{xp^+} - p^- + \frac{\tvec{k}^2 + \lambda^2}{(1-x)p^+}
 \right]}
\\ \nonumber &=& \frac{\pm 2\pi^2 \delta
 \left[
  \frac{\tvec{k}^2}{xp^+} - r^- - \frac{(\tvec{k}-\tvec{r})^2}{(x-\Delta)p^+}
 \right] }{
 \left[
  \frac{\tvec{k}^2}{xp^+}-q^- + \frac{(\tvec{k}-\tvec{q})^2}{(\beta-x)p^+}
 \right]
 \left[
  \frac{\tvec{k}^2}{xp^+} - p^- + \frac{\tvec{k}^2 + \lambda^2}{(1-x)p^+}
 \right]},
\end{eqnarray}
where the sign ambiguity of the $i\epsilon$ components indicates the
presence of a false pole.  Whatever the sign of \eqref{Mcut11}, it is
exactly canceled by the contribution of Res[\ding{174}]
Im[\ding{172}]:
\begin{eqnarray}
\label{Mcut12}
\mathcal{I}_2 &=& \frac{+2\pi \, \mathrm{Im} \left\{ \frac{1}{
  r^- + \frac{(\tvec{k}-\tvec{r})^2-i\epsilon}{(x-\Delta)p^+} - \frac{\tvec{k}^2-i\epsilon}{xp^+}
 } \right\} }{
 \left[
  r^- - q^- + \frac{(\tvec{k}-\tvec{r})^2}{(x-\Delta)p^+} + \frac{(\tvec{k}-\tvec{q})^2}{(\beta-x)p^+}
 \right]
 \left[
  r^- - p^- + \frac{(\tvec{k}-\tvec{r})^2}{(x-\Delta)p^+} + \frac{\tvec{k}^2 + \lambda^2}{(1-x)p^+}
 \right]}
\\ \nonumber &=& \frac{\mp 2\pi^2 \delta
 \left[
  r^- + \frac{(\tvec{k}-\tvec{r})^2}{(x-\Delta)p^+} - \frac{\tvec{k}^2}{xp^+}
 \right] }{
 \left[
  r^- - q^- + \frac{(\tvec{k}-\tvec{r})^2}{(x-\Delta)p^+} + \frac{(\tvec{k}-\tvec{q})^2}{(\beta-x)p^+}
 \right]
 \left[
  r^- - p^- + \frac{(\tvec{k}-\tvec{r})^2}{(x-\Delta)p^+} + \frac{\tvec{k}^2 + \lambda^2}{(1-x)p^+}
 \right]}.
\end{eqnarray}
Thus $\mathcal{I}_1 + \mathcal{I}_2 = 0$, so that proton decay by this
channel is automatically prohibited for Case B.

\begin{itemize}
 \item \underline{ Res[\ding{174}] Im[\ding{173}]: Kinematically Prohibited }
\end{itemize}
This contribution would be
\begin{eqnarray}
\label{Mcut13}
\mathcal{I} &=& \frac{+2\pi^2 \, \delta
 \left[
  r^- - q^- + \frac{(\tvec{k}-\tvec{r})^2}{(x-\Delta)p^+} + \frac{(\tvec{k}-\tvec{q})^2}{(\beta-x)p^+}
 \right] }{
 \left[
  r^- + \frac{(\tvec{k}-\tvec{r})^2}{(x-\Delta)p^+} - \frac{\tvec{k}^2}{xp^+}
 \right]
 \left[
  r^- - p^- + \frac{(\tvec{k}-\tvec{r})^2}{(x-\Delta)p^+} + \frac{\tvec{k}^2 + \lambda^2}{(1-x)p^+}
 \right]}
\\ \nonumber &\propto&
 \delta \bigg[
  \left( (x-\Delta) \tvec{q} + (\beta-x)\tvec{r} - (\beta-\Delta)\tvec{k} \right)^2
 \bigg];
\end{eqnarray}
since the argument is positive definite, this cut is kinematically
prohibited, as it corresponds to a $1 \rightarrow 2$ massless process.

\begin{itemize}
 \item \underline{ Res[\ding{174}] Im[\ding{175}]: Kinematically Prohibited }
\end{itemize}
This contribution would be
\begin{eqnarray}
\label{Mcut14}
\mathcal{I} &=& \frac{+2\pi^2 \, \delta
 \left[
  r^- - p^- + \frac{(\tvec{k}-\tvec{r})^2}{(x-\Delta)p^+} + \frac{\tvec{k}^2 + \lambda^2}{(1-x)p^+}
 \right] }{
 \left[
  r^- + \frac{(\tvec{k}-\tvec{r})^2}{(x-\Delta)p^+} - \frac{\tvec{k}^2}{xp^+}
 \right]
 \left[
  r^- - q^- + \frac{(\tvec{k}-\tvec{r})^2}{(x-\Delta)p^+} + \frac{(\tvec{k}-\tvec{q})^2}{(\beta-x)p^+}
 \right]}
\\ \nonumber &\propto&
 \delta \bigg[
  (x-\Delta)^2 \lambda^2 + \left( (1-x)\tvec{r} - (1-\Delta)\tvec{k} \right)^2
 \bigg];
\end{eqnarray}
since the argument is positive definite, this cut is kinematically
prohibited.

\begin{itemize}
 \item \underline{ Res[\ding{172}] Im[\ding{173}]: Legal Cut }
\end{itemize}
Again, this is the only legal cut of the diagram in Fig.~\ref{MDYB}.
The expression is the same as in \eqref{Mcut10} from Case A, but with
$x > \Delta$.  This means that the delta function
\begin{eqnarray}
  \nonumber
  \delta \left[ x - \left( 1 + 2 \, \frac{\tvec{q} \cdot (\tvec{k} - \tvec{r})}{Q^2} \right) \Delta \right]
\end{eqnarray}
has its singularity within the kinematic window of Case B,
$0<\Delta<x<\beta<1$, if $\tvec{k} \cdot (\tvec{q}-\tvec{r}) > 0$.
Hence we again recover \eqref{MIDY22}, but with validity in the other
half of the $d^2k$ phase space.  Cases A and B thus complement each
other, and we will show that Case C does not make any contribution to
\eqref{MIDY}.

\vspace{0.5cm}

\begin{figure}
\centering
\includegraphics[width=\textwidth]{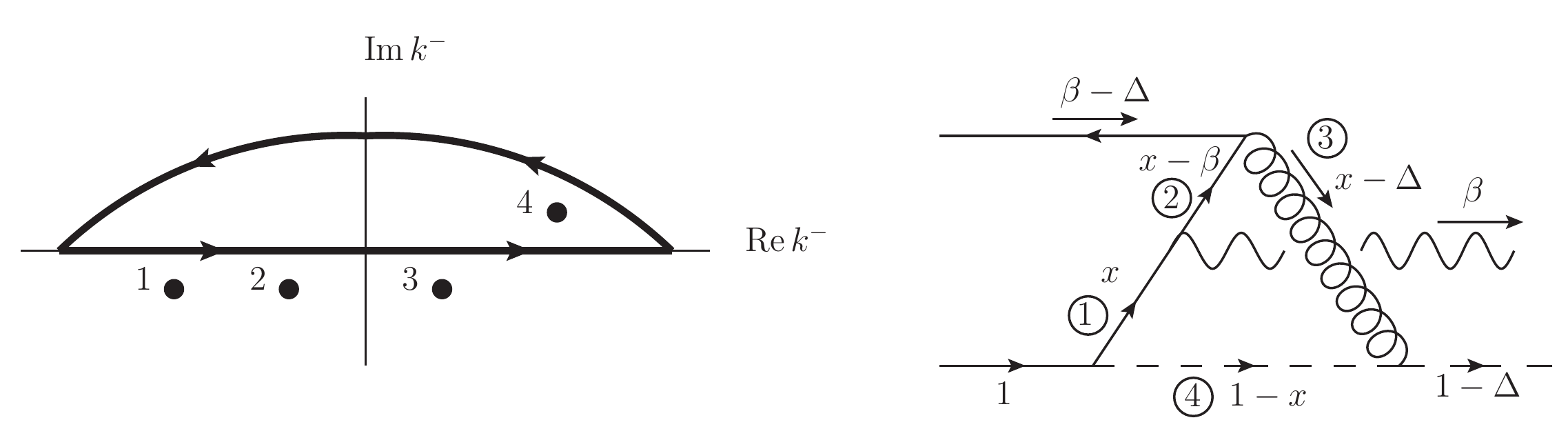}
\caption{\label{MDYC} Illustration of the poles (left) and
  corresponding time-ordered diagram (right) of \eqref{MIDY} for the
  kinematic regime Case C: $0<\Delta<\beta<x<1$.  We choose to close
  the contour in the upper half-plane, enclosing only the pole
  \ding{175}.  Note that the placement of the poles is only schematic,
  indicating the sign of their imaginary part; the placement on the
  real axis has no significance.}
\end{figure}
%
For Case C: $0<\Delta<\beta<x<1$, we choose to close the contour in
the upper half plane, enclosing only the pole \ding{175}, as
illustrated in Fig. \ref{MDYC}.  We demonstrate below that there is no
viable cut for this time-ordering of the process.

\begin{itemize}
 \item \underline{ Res[\ding{175}] Im[\ding{172}]: Proton Decay }
\end{itemize}
This cut is the same as in \eqref{Mcut8} of Case A; closing the
contour in the other direction does not affect the overall sign of the
contribution once the imaginary part is taken, and changing the signs
of $\beta -x$ and $\Delta - x$ does not affect the proton stability
condition.

\begin{itemize}
 \item \underline{ Res[\ding{175}] Im[\ding{174}]: Kinematically Prohibited }
\end{itemize}
This cut is the same as in \eqref{Mcut14} of Case B; closing the
contour in the other direction does not affect the overall sign, and
changing the sign of $\beta-x$ does not affect the argument of the
delta function.  Hence this process is also kinematically forbidden.

\begin{itemize}
 \item \underline{ Res[\ding{175}] Im[\ding{173}]: Kinematically Prohibited }
\end{itemize}
This is the only new cut that requires explicit calculation.  This
contribution would be
\begin{eqnarray}
\label{Mcut15}
\mathcal{I} &=& \frac{+2\pi^2 \, \delta
 \left[
  p^- - q^- - \frac{\tvec{k}^2 + \lambda^2}{(1-x)p^+} - \frac{(\tvec{k}-\tvec{q})^2}{(x-\beta)p^+}
 \right] }{
 \left[
  p^- - \frac{\tvec{k}^2 + \lambda^2}{(1-x)p^+} - \frac{\tvec{k}^2}{xp^+}
 \right]
 \left[
  p^- - r^- - \frac{\tvec{k}^2 + \lambda^2}{(1-x)p^+} - \frac{(\tvec{k}-\tvec{r})^2}{(x-\Delta)p^+}
 \right]}
\\ \nonumber &\propto&
 \delta \bigg[
  -(\beta-\Delta)(x-\beta)(x-\Delta) \lambda^2 - (1-x)(x-\beta)(1-\beta) \left(\tvec{r}-\frac{1-\Delta}
  {1-\beta} \tvec{q} \right)^2
\\ \nonumber &-&
  (1-\Delta)(\beta-\Delta)(1-\beta) \left(\tvec{k} - \frac{1-x}{1-\beta} \tvec{q} \right)^2
 \bigg].
\end{eqnarray}
Since the argument is negative definite, this process is kinematically
forbidden; this cut would not only correspond to proton decay in the
lower half of the diagram in Fig. \ref{MDYC}, but also a kinematically
prohibited $3 \rightarrow 1$ process in the upper half.  Thus we have
shown that there is no viable cut of the diagram for the kinematics of
Case C, and this case makes no contribution to the asymmetry.
Therefore \eqref{MIDY22} gives the complete expression for the
imaginary part and is our final result. 

\eqref{MIDY22} is illustrated in \fig{DIS-DY1} in the text by the
second (shorter) cut putting the quark and anti-quark propagators on
mass shell.



\end{document}